\PassOptionsToPackage{%
  dvipsnames,%
  svgnames,%
  table,%
  usenames,%
  x11names,%
}{xcolor}
\documentclass[%
]{aa}  

\usepackage{graphicx}
\usepackage[varg]{txfonts}
\usepackage{xcolor}
\usepackage{adjustbox}
\usepackage{algorithm2e}
\usepackage{amsfonts}
\usepackage{mathtools}
\usepackage{array}
\usepackage{booktabs}
\usepackage[
  font={sf,footnotesize},%
  labelfont={sf,bf},%
  textfont=sf,%
  skip=0.pt,%
  margin=5pt,%
  format=hang,%
  list=off,%
  figureposition=bottom,%
  tableposition=top%
]{caption}
\usepackage[subnum]{cases}
\usepackage{dsfont}
\usepackage[inline,shortlabels]{enumitem}
\usepackage{float}
\usepackage[%
  pdfauthor={Nuno R. C. Gomes},%
  pdftitle={Prediction of Stellar Rotation Periods Using Tree-based Regression Approaches},%
  pdfsubject={Astronomy, Astrophysics, Machine Learning},%
  pdfkeywords={machine-learning, random-forests, gradient-boosting, xgboost, stellar-rotation-periods, astronomy-astrophysics},%
  pdfusetitle,%
  pdfpagelabels,%
  bookmarks,%
  breaklinks=true,%
  hidelinks,%
  hyperindex,%
  hyperfigures%
]{hyperref} % add links to PDF file
\usepackage{bookmark}
\usepackage[]{hyphenat} % [htt]
\usepackage{ifthen}
\usepackage{interval}
\usepackage{lscape}
\usepackage{multicol}
\usepackage{natbib}
\usepackage[%
  detect-family,%
  detect-weight%
]{siunitx}
\usepackage{subcaption}
\usepackage{tabulary}
\usepackage{tikz}
\usepackage{etoolbox}
\usepackage{cleveref}
\definecolor{darkpastelgreen}{rgb}{0.01, 0.75, 0.24}
\definecolor{deepsaffron}{rgb}{1.0, 0.6, 0.2}
\definecolor{tomato}{rgb}{1.0, 0.39, 0.28}
\graphicspath{{./imgs/}, {./figs}, {./figures/}}
\newif\ifshowchanges % define boolean switch 'showchanges'
\newif\ifnost % define boolean switch 'nostrikethrough'
\newcommand{\added}[1]{%
  \ifshowchanges%
    {\color{RoyalBlue} #1}%
  \else%
    \ifnost%
      {\color{RoyalBlue}\bf#1}%
    \else%
      #1%
    \fi%
  \fi%
}%
\newcommand{\deleted}[2][red]{%
  \ifshowchanges{%
    \begin{tikzpicture}[baseline=(text.base)]%
      \node[inner sep=0pt, outer sep=0pt] (text) {#2};%
      \draw[#1, line width=1.0pt] (text.west) -- (text.east);%
    \end{tikzpicture}%
  }\else{}%
  \fi%
}
\newcommand{\eg}{\textit{e.g.}, {}}

\newcommand{\ie}{\textit{i.e.}, {}}

\newcommand{\replaced}[3][red]{%
  \ifshowchanges
    {\color{OrangeRed}{#2}}%
    {%
      \begin{tikzpicture}[baseline=(text.base)]
        \node[inner sep=0pt, outer sep=0pt] (text) {#3};
        \draw[#1, line width=1.0pt] (text.west) -- (text.east);
      \end{tikzpicture}%
    }%
  \else%
    \ifnost%
      {\color{OrangeRed}\bf#2}%
    \else%
      #2%
    \fi%
  \fi%
}%
\newcommand{\repdel}[3][OrangeRed]{%
  \ifshowchanges{%
    \textcolor{#1}{#2}% usage: \repdel[col]{new text}{old text}
  }%
  \else{#2}%
  \fi
}%
\newcommand{\rlang}{\textsf{R}}
\newcommand{\vs}{\textit{vs}.\ {}}

\newcommand{\xgb}{\textbf{\textsf{xgboost}}}

\begin{document} 
\title{Predicting Stellar Rotation Periods Using XGBoost}
\author{
  Nuno R. C. Gomes, \orcid{0000-0003-4864-9530}\inst{1, 2, 3} \thanks{E-mail: ngomes@ieec.cat}
  Fabio Del Sordo, \orcid{0000-0001-9268-4849}\inst{1, 4, 5} \thanks{E-mail: delsordo@ice.csic.es}
  Luís Torgo\inst{2, 3, 6}\thanks{Deceased on 11 April 2023}
}          
\date{Received -- / Accepted --}
\institute{
  Institut d'Estudis Espacials de Catalunya (IEEC), 08860 Castelldefels (Barcelona), Spain
  \and
  Departamento de Ciências de Computadores, Faculdade de Ciências, Universidade do Porto, rua do Campo Alegre s/n, 4169-007 Porto, Portugal
  \and
  Faculty of Computer Science, Dalhousie University, 6050 University Avenue, PO BOX 15000, Halifax, NS B3H 4R2, Canada
  \and
  Institute of Space Sciences (ICE-CSIC), Campus UAB, Carrer de Can Magrans s/n, 08193, Barcelona, Spain
  \and
  INAF, Osservatorio Astrofisico di Catania, via Santa Sofia, 78 Catania, Italy
  \and
  LIAAD INESC Tec, INESC, Campus da FEUP, Rua Dr.\ Roberto Frias, 4200-465 Porto, Portugal
}

%%% fix \qty in older TeXLive compiler %%%%%%%%%%%
\let\qty\SI
\let\qtylist\SIlist
\let\qtyrange\SIrange
%%%%%%%%%%%

% \abstract{}{}{}{}{} 
% five {} token are mandatory

\abstract
{
  The estimation of rotation periods of stars is a key problem in stellar astrophysics.
  Given the large amount of data available from ground-based and space-based telescopes, there is a growing interest in finding reliable methods to quickly and automatically estimate stellar rotation periods with accuracy and precision.
}
{
  This work aims to develop a computationally inexpensive approach, based on machine learning techniques, to accurately predict thousands of stellar rotation periods.
}
{
  The innovation in our approach is the use of the XGBoost algorithm to predict the rotation periods of \textit{Kepler} targets by means of regression analysis.
  Therefore,
  we focused on building a robust supervised machine learning model to predict surface stellar rotation periods from structured data sets built from the \textit{Kepler} catalogue of K and M stars.
  We analysed the set of independent variables extracted from \textit{Kepler} light curves and investigated the relationships between them and the ground truth.
}
{
  % NG
  Using the extreme gradient boosting method, we obtained a minimal set of variables that can be used to build machine learning models for predicting stellar rotation periods.
  Our models are validated by predicting the rotation periods of about \num{2900} stars.
  The results are compatible with those obtained by classical techniques and comparable to those obtained by other recent machine learning approaches, with the advantage of using much fewer predictors.
  Restricting the analysis to stars with rotation periods of less than 45 days, our models are on average \deleted{95 to}\replaced{\qty{96}{\percent}}{\qty{98}{\percent}} correct.
}
{
  We have developed an innovative approach, based on a machine learning method, to accurately fit the rotation periods of stars.
  Based on the results of this study, we conclude that the best models generated by the proposed methodology are competitive with the state-of-the-art approaches, with the advantage of being computationally cheaper, easy to train, and relying on small sets of predictors.
}


\maketitle
%
%-------------------------------------------------------------------

\section{Introduction}
\label{sec:intro}
Measuring how stars rotate is an essential part of stellar astrophysics. 
One of the main methods for quantifying the rotation periods of stars is to analyse their light curves and look for modulations that can be related to the stellar rotation rate.
Solar-type stars, \ie low-mass stars with convective outer layers, are known to develop spots on their surfaces.
The origin of these spots is related to stellar magnetism \citep[\eg][]{brun2017magnetism}.
Similarly, high-mass \citep[\eg][]{cantiello2011magnetic} and fully convective low-mass stars \citep[\eg][]{bouvier1989spots,damasso2020low,bicz2022starspot} also exhibit magnetic spots on their surfaces.
Such spots can induce a modulation in stellar light curves which, in principle, allows the determination of both stellar rotation periods and long-term stellar magnetic activity cycles \citep[\eg][]{strassmeier2009starspots}.
The rotation period of a star is essential for the understanding of the transport of stellar angular momentum, a process that is still poorly understood \citep{aerts2019angular}, and is important for the correct estimation of the age of stars \citep{eggenberger2009effects}.
The latter can be very important for the characterisation of planetary systems \citep{huber2016k2}.
The role of stellar rotation in driving stellar dynamos and determining magnetic cycles is also still much debated \citep[e.g.][]{bonanno2022origin}.
Any theory on this subject requires an accurate calculation of rotation periods for all types of stars.

The accuracy and precision with which stellar rotation is measured are crucial for the study of stellar evolution.
The rotation period of a star correlates with its age: solar-type stars are known to spin down during their main-sequence evolution, and so the rotation period of young solar-like stars can be used to constrain stellar ages using \textit{gyrochronology relations} \citep{barnes2003rotational,skumanich1972time,garcia2014rotation,messina2021constraining,messina2022gyrochronological}.
However, for stars older than the Sun, gyrochronological ages do not agree with asteroseismic ages \citep{vansaders2016weakened,hall2021weakened} and those inferred from velocity dispersion \citep{angus2020exploring}.
Such a discrepancy may then be solved only by developing new methods for accurately measuring stellar rotation.

The vast amount of astronomical photometric data released over the last three decades has recently motivated the use of machine learning (ML) techniques to process and analyse it.
The advent of new large-scale sky surveys and the need to process large numbers of targets simultaneously make manual handling of astrophysical data impractical, and the use of artificial intelligence (AI) techniques is becoming increasingly popular \citep[\eg][]{pichara2016meta,biehl2018machine}.
A stellar light curve is nothing more than a \textit{time series} of photometric data from a star, \ie a sequence of stellar surface fluxes collected at successive points in time.
The last decade has seen the emergence of many observations that provide high-quality, long-term and near-continuous stellar photometric data.
Examples include the \textit{Kepler} space observatory \citep{borucki2009kepler} and the reborn \textit{Kepler} K2 mission \citep{howell2014k2}, which together have observed more than half a million stars \citep{huber2016k2},\footnote{
  \url{https://exoplanets.nasa.gov/resources/2192/nasas-kepler-mission-by-the-numbers/}
} and the Transiting Exoplanet Survey Satellite \citep[TESS,][]{ricker2015tess}, which has collected light curves with time spans of 25 days to one year for tens of millions of stars.
Such a large number of observations requires automated procedures to process and extract information from them, and machine learning methods can be used to do this.

The first step in choosing an ML technique is to select the data from available sources and study them carefully.
Typically, two approaches can be adopted: using the photometric time series data directly as input, or first converting the light curves into structured data that can be represented by a set of variables or features in tabular form.
Machine learning models can be trained from these two types of data (unstructured and structured), automating processes that would otherwise be tedious or require too much human effort.

The first case---unstructured data---is usually tackled with special types of \textit{artificial neural networks} \citep[ANN,][]{haykin2009neural}, reinforcement algorithms that require little or no pre-processing of the data.
\citet{blancato2022data} used a \textit{deep learning} (DL) approach and applied \textit{convolutional neural networks} (CNN) to predict stellar properties from \textit{Kepler} data.
\citet{claytor2022recovery} applied a CNN to synthetic light curves to infer stellar rotation periods\added{, %
  and then estimated the rotation periods of \num{7245} main-sequence TESS stars, with periods up to 35 days for G, K dwarfs and up to 80 days for M dwarfs \citep{claytor2024tess}.
  This was achieved using a CNN, which allowed them to remove light curve systematics related to the telescope's 13.7-day orbit.
}
However, training neural networks typically requires large computational resources.

The second scenario---using structured data---is solved by resorting to algorithms that can use tabular data to perform unsupervised (clustering) and supervised (classification and regression) tasks.
\added{%
  \citet{lu2020astraea} developed a model that predicts the rotation periods of TESS stars with an uncertainty of \qty{55}{\percent}, by training a \textit{random forest} \citep[RF,][]{breiman2001random} regressor on \textit{Kepler} data.
  The most important features for predictions in their model were the range of variability in the light curve, the effective temperature, the \textit{Gaia} colour, the luminosity, and the brightness variation on timescales of \qty{8}{\hour} or less.
}
\citet{breton2021rooster} applied \replaced{RFs}{\textit{random forests}}\deleted{\citep{breiman2001random}} to tabular \textit{Kepler} data to produce three ML classifiers that can detect the presence of rotational modulations in the data, flag close binary or classical pulsator candidates, and to provide rotation periods of K and M stars from the \textit{Kepler} catalogue \added{\citep{brown2011kepler}}.
The method was then applied to F and G main-sequence and cool subgiant \textit{Kepler} stars by \citet{santos2021surface}.
\citet{breton2021rooster} used 159 different inputs to train the classifiers: rotation periods, stellar parameters such as mass, effective temperature, and surface gravity (just to name a few), and complementary variables obtained from wavelet analysis, the autocorrelation function of light curves and the composite spectrum.
They claim accuracies of \qty{95.3}{\percent} when willing to accept errors within \qty{10}{\percent} of the reference value, and of \qty{99.5}{\percent} after visual inspection of \qty{25.2}{\percent} of the observations.
\citet{breton2021rooster} used a classification approach, \ie their algorithm was trained to select the most reliable rotation period among these candidate values.
A particularity of their work is therefore that they used rotation periods as input variables for training their models. 
We expected these features to be highly correlated with the response or target variable (the rotation periods considered as ground truth), and so we decided to
\begin{enumerate*}[label=(\roman*), font=\itshape]
  \item explore their data set and determine what is this level of correlation, and
  \item try to train ML models without these rotation period input variables and compare the performance of our models with theirs.
\end{enumerate*}

In this paper, we focus on the prediction of rotation periods of K and M stars from the \textit{Kepler} catalogue \citep{borucki2009kepler,borucki2010kepler}, using an ML method.
We follow a regression-based approach, \ie we propose to predict the rotation period of several targets, rather than selecting the best candidate among some previously computed values, as in a classification problem. 
We address the selection of suitable data, the size of the data set, the optimisation of the ML model parameters, and the training of the latter.
Time and computational resources were also important constraints in the development of this project, especially during the learning phase of the models.

The paper is structured as follows: in \cref{sec:materials,sec:methods}, we describe the materials---the data---and the methods used for the experiments;
\cref{sec:experimental-design} is a discussion of our experimental design;
in \cref{sec:results}, we present the results of our contributions by applying a supervised ML approach to predict the rotation periods of stars;
\cref{sec:discussion} is dedicated to the discussion of the results, and we summarise our contributions in \cref{sec:conclusions}.
%%%%%%%%%%%%%%%%%%%%%%%%%%%%%%%%%%%%%%%%%%%%%%%%%%%%%%%%%%%%%%%%%%%%%%%%%%%%%%%%

\section{Materials}
\label{sec:materials}
The main objective of this project is to develop robust yet computationally inexpensive supervised ML models from tabular astronomical data for the prediction of stellar rotation periods of K and M stars from the \textit{Kepler} catalogue.

We used sets of structured, tabular data, containing measurements for the \textit{features} or \textit{predictors} and for the \textit{response} or \textit{target} variable.
Computationally, these tabular data have been organised into \textit{data frames}, where the columns correspond to the variables (predictors plus response) and the rows correspond to the observations.
Each of the rows, consisting of measurements for one star, will be referred to as an \textit{observation}, an \textit{instance}, a \textit{case}, or an \textit{object}.

These data sets were divided into \textit{training} and \textit{testing} sets.
The former were used to build prediction \textit{models} or \textit{learners}, which in turn were used to predict the rotation period of unseen stars, provided they were input as structured data sets similar to the training set, \ie in tabular form, containing at least some of its features.
The testing sets acted as the never-before-seen objects, with the advantage of containing the outcome (not given to the model), which could be used to assess the predictive performance and quality of the models by comparing the predicted values with the true outcomes or \textit{ground truth}.
We consider a good model to be one that accurately predicts the response.

The following sections describe the specific materials and methods used to build the models and assess their performance.\footnote{
  The \rlang\ programming language \citep{rcoreteam2023r} was used for all code and data analyses.
}
%%%-----------------------------------------------------------------------------

\subsection{Description of the Data}
\label{sec:data}
We focused on real data, including features extracted from the light curves, and ``standard'' predictors, \ie variables that are commonly obtained directly or indirectly from astronomical observations.

We started by analysing structured data from the \textit{Kepler} catalogue of K and M dwarf stars, already in tabular form, published by \citet[][]{santos2019surface} and \citet[][]{breton2021rooster}, whose targets were selected from the \textit{Kepler} Stellar Properties Catalogue for Data Release 25 \citep{mathur2017revised}.
Both catalogues---hereafter referred to as S19 and B21, respectively---contain all the predictors and targets considered in our work: B21 was used as the source for most of the predictors; S19 was used specifically to extract the rotation periods and features obtained directly from \textit{Kepler} observations.
The latter are commonly known as ``stellar parameters'' in the astronomical community---examples include the mass of the star and its effective temperature, to name but a few.
However, we will not use the word ``parameter'' in this context to avoid confusion with \textit{model parameters}.
Instead, we will use the terminology of \textit{astrophysical variables} when referring to them.

B21 is available in tabular form, without any particular clustering or classification of the variables.
However, we decided to group the predictors according to their nature and/or the method by which they were obtained.
We have identified six groups, with the following characteristics:\footnote{%
  Words in brackets, in \texttt{typewriter font}, refer to variable names as they appear in \cref{tab:app-variables} and in the data frames.
}
\begin{enumerate}\label{page:variable-families}
  \item \textit{Astrophysical} (Astro, \texttt{astro}) --- predictors related to the physical properties of the stars, derived directly or indirectly from the observation, to wit:
    \begin{itemize}
      \item effective temperature, $T_\text{eff}$ (\texttt{teff}), and its corresponding upper and lower errors, $T_\text{eff}^\text{err, up}$ (\texttt{teff\_eup}) and $T_\text{eff}^\text{err, low}$ (\texttt{teff\_elo}) respectively;
      \item the logarithm of the surface gravity, $\log g$ (\texttt{logg}) and its upper and lower error limits, $\log g^\text{err, up}$ (\texttt{logg\_eup}) and $\log g^\text{err, low}$ (\texttt{logg\_elo});
      \item the mass of the star, $M$ (\texttt{m}), and its upper and lower errors, $M^\text{err, up}$ (\texttt{m\_eup}) and $M^\text{err, low}$ (\texttt{m\_elo});
      \item the magnitude from the \textit{Kepler} input catalogue, $K_\text{p}$ (\texttt{kepmag}); and
      \item the Flicker in Power or \textit{FliPer} values, $F_{0.7}$ (\texttt{f\_07}), $F_7$ (\texttt{f\_7}), $F_{20}$ (\texttt{f\_20}), and $F_{50}$ (\texttt{f\_50}), respectively corresponding to cut-off frequencies of \qtylist[list-units= single]{0.7;7;20;50}{\micro\hertz}.%\footnote{
    \end{itemize}
  \item \textit{Time Series} (TS, \texttt{ts}) --- quantities that are related to the properties of the time series:
  \begin{itemize}
    \item the length of the light curve in days, $l$ (\texttt{length});
    \item the start and end times of the light curve, $S_\text{t}$ (\texttt{start\_time}) and $E_\text{t}$ (\texttt{end\_time}) respectively;
    % \item the bad quarter flag, $Q_\text{bad}$ (\texttt{bad\_q\_flag});
    \item the number of bad quarters in the light curve, $nQ_\text{bad}$ (\texttt{n\_bad\_q});
    \item the photometric activity proxy or index, $S_\text{ph}$ (\texttt{sph}), and its error, $S_\text{ph}^\text{err}$ (\texttt{sph\_e}), according to \cite{santos2019surface};
    \item the photometric activity proxy computed from the autocorrelation function (ACF) method to obtain the rotation period, $S_\text{ph}^\text{ACF}$ (\texttt{sph\_acf\_xx}), for each of the $xx$-day filters, where $xx \in \{20,\, 55,\, 80\}$, as provided by \cite{breton2021rooster}, and its corresponding error, $S_\text{ph}^\text{ACF, err}$ (\texttt{sph\_acf\_err\_xx});\footnote{
      In \citet{breton2021rooster}, $S_\text{ph}$ is divided into several values, each corresponding to the process from which it was obtained (ACF, CS, or GWPS) and to the filter applied to the light curves (20, 55, or 80 days).
    }
    \item the height of the $P_\text{ACF}$ (\ie the period of the highest peak in the ACF at a lag greater than zero), $G_\text{ACF}$ (\texttt{g\_acf\_xx}), for each of the $xx$-day filters; and
    \item the mean difference between the heights of $P_\text{ACF}$ at the two local minima on either side of $P_\text{ACF}$, $H_\text{ACF}$ (\texttt{h\_acf\_xx}), for each of the above filters.
  \end{itemize}
  \item \textit{Global Wavelet Power Spectrum} (GWPS, \texttt{gwps}) --- quantities obtained from a time-period analysis of the light curve using a wavelet decomposition:
  \begin{itemize}
    \item the amplitude (\texttt{gwps\_gauss\_1\_j\_xx}), central period (\texttt{gwps\_gauss\_2\_j\_xx}), and standard deviation (\texttt{gwps\_gauss\_3\_j\_xx}) of the $j^\text{th}$ Gaussian fitted to the GWPS with the $xx$-day filter;
    \item the mean level of noise of the Gaussian functions fitted to the GWPS for the $xx$-day filter (\texttt{gwps\_noise\_xx});
    \item the $\chi^2$ of the fit of the Gaussian function to GWPS for the $xx$-day filter (\texttt{gwps\_chiq\_xx});
    \item the number of Gaussian functions fitted to each GWPS for the $xx$-day filter (\texttt{gwps\_n\_fit\_xx}); and
    \item the photmetric proxy computed from the GWPS method, $S_\text{ph}^\text{GWPS}$ (\texttt{sph\_gwps\_xx}), as provided by \citet{breton2021rooster}, and its corresponding error, $S_\text{ph}^\text{GWPS, err}$ (\texttt{sph\_gwps\_err\_xx}).
  \end{itemize}
  \item \textit{Composite Spectrum} (CS, \texttt{cs}) --- variables obtained from the product of the normalised GWPS with the ACF:%\footnote{
  \begin{itemize}
    \item the amplitude of the $P_\text{CS}$ (the period of the fitted Gaussian with the highest amplitude), $H_\text{CS}$ (\texttt{h\_cs\_xx}), for the $xx$-day filter;
    \item the average noise level of the Gaussian functions fitted to the CS, $\mathit{cs}_\text{noise}$ (\texttt{cs\_noise\_xx}), for the $xx$-day filter;
    \item the $\chi^2$ of the fit of the Gaussian function to the CS for the $xx$-day filter (\texttt{cs\_chiq\_xx});
    \item the photometric proxy computed by the CS method, $S_\text{ph}^\text{CS}$ (\texttt{sph\_cs\_xx}), and its error, $S_\text{ph}^\text{CS, err}$ (\texttt{sph\_cs\_err\_xx}) for the $xx$-day filter; and
    \item the amplitude (\texttt{cs\_gauss\_1\_j\_xx}), central period (\texttt{cs\_gauss\_2\_j\_xx}), and the standard deviation (\texttt{cs\_gauss\_3\_j\_xx}) of the $j^{th}$ Gaussian fitted to the CS with the $xx$-day filter, for $j\in\{1,\, 2,\, \ldots,\, 6\}$.
  \end{itemize}
  \item \textit{Rotation Periods} (Prot, \texttt{prot}) --- rotation periods obtained by combining the ACF, CS, and GWPS time-period analysis methods with the KEPSEISMIC light curves, for the $xx$-day filters: $P_\text{rot}^\text{ACF}$ (\texttt{prot\_acf\_xx}), $P_\text{rot}^\text{CS}$ (\texttt{prot\_cs\_xx}), and $P_\text{rot}^\text{GWPS}$ (\texttt{prot\_gwps\_xx}).\footnote{
    The KEPSEISMIC time series are available from the Mikulski Archive for Space Telescopes (MAST), via the link \url{https://dx.doi.org/10.17090/t9-mrpw-gc07}.
  }
\end{enumerate}
The list of 170 explanatory variables, as they appear in the final data set after exploratory data analysis, cleaning, and engineering, is referenced in \cref{tab:app-variables}.
In total, there are 14 Astro, 18 TS, 66 GWPS, 63 CS, and nine Prot predictors. 

\begin{sidewaystable*}
  \caption{Names of the 170 predictors, as they appear in the original data set, and their description. Indicated are the group to which they belong, and the effective number of variables associated to each of them ($N$).}
  \label{tab:app-variables} 
  \centering
  \small
  \begin{tabulary}{\linewidth}{lLcLc}
    \toprule
    Variable & Description & Group & Observations & $N$ \\
    \midrule
    \texttt{teff} & $T_\text{eff}$, effective temperature of the star & astro & & 1 \\
    \texttt{teff\_eup}, \texttt{teff\_elo} & upper and lower errors for $T_\text{eff}$ & astro & & 2 \\
    \texttt{logg} & $\log g$, logarithm of the surface gravity & astro & & 1 \\
    \texttt{logg\_eup}, \texttt{logg\_elo} & upper and lower errors of $\log g$ & astro & & 2 \\
    \texttt{m} & $M$, the mass of the star, in solar masses & astro & & 1 \\
    \texttt{m\_eup}, \texttt{m\_elo} & upper and lower errors for $M$ & astro & & 2 \\
    \texttt{f\_07}, \texttt{f\_7}, \texttt{f\_20}, \texttt{f\_50} & FliPer values for $\nu_\text{C} = \SIlist[list-units= single]{0.7;7;20;50}{\micro\hertz}$ & astro & & 4 \\
    \texttt{kepmag} & \textit{Kepler} magnitude from the \textit{Kepler} input catalogue & astro & parameter linked to the quality of the stellar target & 1 \\
    \addlinespace
    \texttt{length} & length of the light curve, in days & ts & parameter linked to the quality of the acquired light curve & 1 \\
    \texttt{n\_bad\_q} & number of bad quarters in the light curve & ts & parameter linked to the quality of the acquired light curve & 1 \\
    \texttt{start\_time}, \texttt{end\_time} & starting and ending time of the light curve & ts & parameter linked to the quality of the acquired light curve & 2 \\
    \texttt{sph\_acf\_xx}, \texttt{sph\_acf\_err\_xx} & $S_\text{ph}$, photometric activity proxy, and its error, computed on the ACF for the $xx$-day filter & ts & mean of standard deviations over light curve segments of $5 \times P_\text{rot}^\text{ACF}$ & 6 \\
    \texttt{sph}, \texttt{sph\_e} & $S_\text{ph}$, photometric activity proxy & ts & it is a magnetic activity proxy & 2 \\
    \texttt{g\_acf\_xx} & height of $P_\text{ACF}$ for the $xx$-day filter & ts & $P_\text{ACF}$ is the period of the highest peak in the ACF at a lag greater than zero; $G_\text{ACF}$ is a control parameter & 3 \\
    \texttt{h\_acf\_xx} & mean difference between height of $P_\text{ACF}$ and the two local minima on both its sides for the $xx$-day filter & ts & $H_\text{ACF}$ is a control parameter & 3 \\
    \addlinespace
    \texttt{cs\_noise\_xx} & mean level of noise of the Gaussian functions fitted to CS for the $xx$-day filter & cs & this is the period uncertainty; the level of noise corresponds to the half width at half maximum (HWHM) & 3 \\
    \texttt{cs\_chiq\_xx} & $\chi^2$ of the fit of the Gaussian function on CS for the $xx$-day filter & cs & & 3 \\
    \texttt{cs\_n\_fit\_xx} & number of Gaussian function fitted to each CS for the $xx$-day filter & cs & & 3 \\
    \texttt{h\_cs\_xx} & amplitude of $P_\text{CS}$ for the $xx$-day filter & cs & $P_\text{CS}$ is obtained as the period of the fitted Gaussian of highest amplitude; it is a control parameter & 3 \\
    \texttt{sph\_cs\_xx}, \texttt{sph\_cs\_err\_xx} & $S_\text{ph}$ and its error computed on CS for the $xx$-day filter & cs & mean standard deviations over light curve segments of $5\times P_\text{rot}^\text{CS}$ & 6 \\
    \texttt{cs\_gauss\_i\_j\_xx} & amplitude ($i=1$), central period ($i=2$), and standard deviation ($i=3$) of the $j^\text{th}~(j=1, 2, \ldots5)$ Gaussian fitted in the CS with the $xx$-day filter & cs & CS is the product between the normalised GWPS and the ACF & 45 \\
    \addlinespace
    \texttt{gwps\_noise\_xx} & mean level of noise of the Gaussian functions fitted to GWPS for the $xx$-day filter & gwps & this is the period uncertainty; the level of noise is given by HWHM & 3 \\
    \texttt{gwps\_n\_fit\_xx} & number of Gaussian functions fitted to each GWPS for the $xx$-day filter & gwps & & 3 \\
    \texttt{gwps\_gauss\_i\_j\_xx} & amplitude ($i= 1$), central period ($i=2$), and standard deviation ($i=3$) of the $j^\text{th}~(j=1, 2, \ldots 6)$ Gaussian fitted in GWPS with the $xx$-day filter & gwps & & 54 \\
    \texttt{sph\_gwps\_xx}, \texttt{sph\_gwps\_err\_xx} & $S_\text{ph}$ and its error computed on GWPS for the $xx$-day filter & gwps & mean standard deviation over light curve segments of $5\times P_\text{rot}^\text{GWPS}$ & 6 \\
    \addlinespace
    \texttt{prot\_acf\_xx} & rotation period extracted from the ACF analysis for the $xx$-day filter & prot & & 3 \\
    \texttt{prot\_cs\_xx} & rotation period extracted from the CS analysis for the $xx$-day filter & prot & & 3 \\
    \texttt{prot\_gwps\_xx} & rotation period extracted from the GWPS analysis for the $xx$-day filter & prot & & 3 \\
    \bottomrule
  \end{tabulary}
\end{sidewaystable*}

We chose to use previously published real-world data because, on the one hand, they are readily available;
and, on the other hand, because we wanted to compare distinct machine learning approaches using different sets of predictors---thus analysing how much the predictions are affected by the variables closely related to the method used to estimate the reference rotation periods---and to compare the results with the state of the art---comparisons that can only be made fairly if the data sets are built from the same reference set of observations.

In the following section, we will describe the steps we applied to S19 and B21 to build a master data set and several subsets which we used to build the ML models.
%%%-----------------------------------------------------------------------------

\subsection{Data Engineering}
\label{sec:data-engineering}
S19 and B21 were cleaned and merged to create a master data set, from which several subsets were generated.
During the cleaning process,
\begin{enumerate*}[label=(\roman*), font=\itshape]
  \item columns containing only null values were dropped,
  \item rows with less than \SI{5}{\percent} of non-null predictors were removed,
  \item rows containing extreme standard deviation values were removed,
  \item very extreme values of variables other than standard deviations were flagged as \texttt{NaN} and subsequently imputed,
  \item rows corresponding to targets with rotation errors greater than \qty{20}{\percent} of the respective rotation period were removed,
  \item flag-type variables from S19 and B21 were dropped,
  \item all variable names were converted to lowercase, and 
  \item some names were changed, for clarity.
\end{enumerate*}

In the merging process, we selected all stars that were present simultaneously in both data sets.
As explained in detail in \cref{sec:problem-formulation}, we chose to use the XGBoost (XGB) version of the tree-based ensemble approach known as \textit{gradient boosting} (GB) to train our models. 
Missing values were imputed using a random forest (RF) with 500 trees and a sample fraction of 0.5.
Following data imputation, we decided to create a parallel data set with standardised predictors for further analysis.
We note that because we used a tree-based (as opposed to a distance-based) ensemble method, normalisation of the predictors would not be strictly required \citep{murphy2012machine,chen2016xgboost}.
Nevertheless, since XGBoost uses a gradient descent algorithm, we decided to use the normalised version of the data set to create the subsets from which the XGB models would be built, in order to optimise the computations during the training phase.

Several features, mainly in the CS and GWPS groups (recall the group in variables in the previous section), contained values with extremely large or infinite amplitudes.
They are derived variables, corresponding to some of the amplitudes, central periods, standard deviations, and mean noise levels of the Gaussian functions fitted to the CS and GWPS.
Their extreme values may have different origins, such as instrumental errors, and possibly they may have been taken into account when they were created.
In principle, the presence of outliers would not be a problem, as there are no restrictions on extreme points in GB, and XGB can handle missing values.
However, as we are validating a methodology and want to leave open the possibility of comparing the performance of XGBoost with other methods that cannot handle missing values (such as some implementations of RFs), we chose to flag these outliers as missing values and subsequently impute them.
We note that the main data set still contains large values from which the models could learn. 

After the aforementioned transformations, we ended up with a data set containing \num{14913} observations and 170 predictors.
This is the full or master set, labelled \textit{Data Set 0} (DS0).

Several subsets were created from the full data set, each containing a different number of predictors and/or instances:

\begin{enumerate}\label{page:subsets}
  \item The first subset was obtained by removing the nine rotation period variables (Prot group), resulting in a data set of 163 variables. This is referred to as \textit{Data Set 1} (DS1).
  \item A second subset was created by removing from DS1 redundant variables, correlated between them, resulting in a data set with 152 predictors --- \textit{Data Set 2} (DS2).
  \item The third subset was obtained from DS2, by removing stars with surface rotation periods greater than \qty{45}{\day}, leaving a data set of \num{14356} rows --- this is \textit{Data Set 3} (DS3).
  \item The fourth subset is obtained by removing from DS3 stars with surface rotation periods less than \qty{7}{\day}, resulting in a data set of \num{13602} rows --- \textit{Data Set 4} (DS4).
  \item A fifth subset was built on top of DS3, using the \replaced{twenty-eight}{thirty-four} most important variables as assessed by the previous models --- \textit{Data Set 5} (DS5).
  \item A sixth subset was obtained from DS4, using the \replaced{twenty-eight}{thirty-four} most important variables as scored by the models built with DS0 to DS4 --- \textit{Data Set 6} (DS6).
  \item A final subset was created from DS1, where we used the same \replaced{twenty-eight}{thirty-four} most important variables used to build DS5 and DS6 --- \textit{Data Set 7}  (DS7).
\end{enumerate}
 
DS2 was constructed by performing a simple univariate filtering of the features, where the predictors were cross-validated with the response variable in order to remove redundant or uninformative variables from the learning process.
The analysis of correlations between variables is detailed in \cref{sec:stats}.

DS3 was built because of \textit{Kepler}'s 90-day quarters \citep{mullally2020mast}, to ensure at least two full cycles of observations per object, \replaced{thus preventing}{and to avoid} any problems \replaced{that may arise}{arising} from stitching together data from different quarters \added{and other potential long-term issues related to the telescope, such as those related to \textit{Kepler}'s orbital period and its harmonics}.
Although \cite{breton2021rooster} have identified and distinguished rotating stars from other types of objects, we also removed stars with surface rotation periods less than \qty{7}{\day}, because these targets can easily be mistaken for close-in binary systems\added{,} \deleted{or} classical pulsators, \added{or other false positives,} whose signals may mimic stellar objects manifesting surface rotation, and we wanted to check how the model would behave by removing these potentially misclassified targets.
By including only stars with rotation periods between \qty{7}{\day} and \qty{45}{\day}, as we did in DS4, we expected to improve the predictive performance of the models.
We ended up with \num{13602} stellar objects, a reduction of about \qty{9}{\percent} compared to the DS0, DS1, and DS2 data sets.

We also selected the top-20 most important variables in the models trained with the data sets DS0 to DS4, and took their union.
This resulted in a set of \replaced{28}{34} predictors, which we used to build DS5\added{,}\deleted{and} DS6\added{, and DS7}.
These \replaced{three}{two} data sets contain only the most important features identified during the learning of the previous XGB models.
%%%-----------------------------------------------------------------------------

\subsection{Statistical Analysis of Relevant Variables}
\label{sec:stats}
\begin{figure*}[!ht]
  \centering
  \begin{subfigure}[b]{0.95\textwidth}
    \centering
    \includegraphics[width=\textwidth]{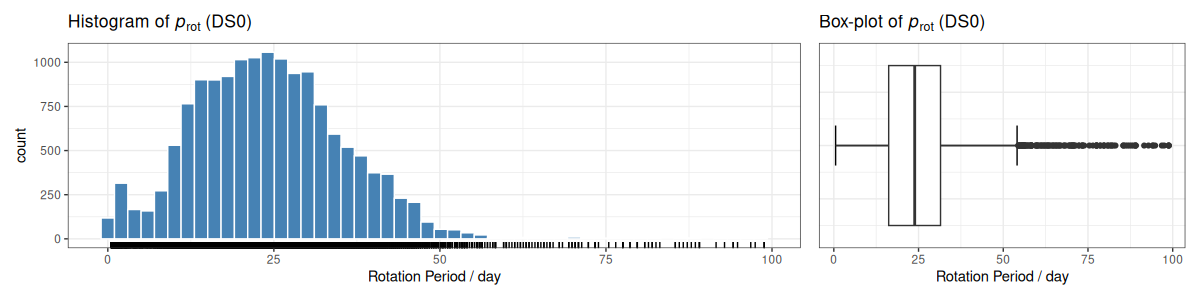}
  \end{subfigure}
  \vspace{5mm}
  \begin{subfigure}[b]{0.95\textwidth}
    \centering
    \includegraphics[width=\textwidth]{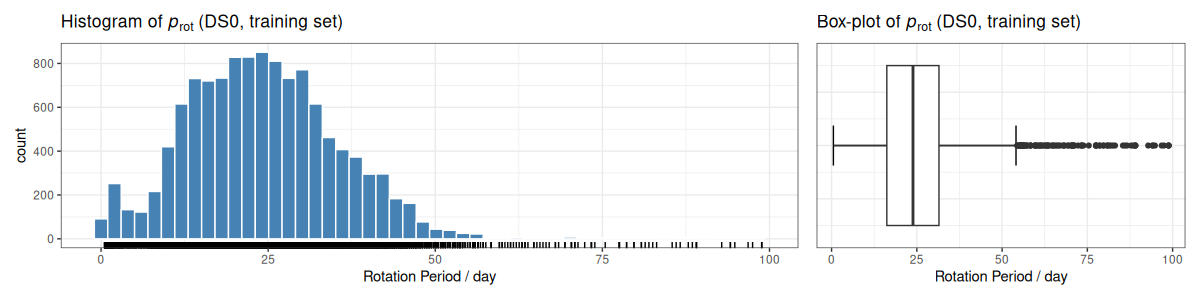}
  \end{subfigure}
  \caption{
    Histograms (left panels) and box-plots (right panels) of the target variable, $p_\text{rot}$, on DS0 (top row) and on the training set generated from it (bottom row).
    The distributions are very similar to each other, right skewed, with medians equal to \qty{23.87}{\day}, minima and maxima of \qty{0.53}{\day} and \qty{98.83}{\day}, respectively.
    Both have several outliers.
  }
  \label{fig:prot-hist-bp-ds0}
\end{figure*}
\begin{figure}[ht]
  \centering
  \begin{subfigure}[b]{0.45\textwidth}
    \centering
    \includegraphics[width=\textwidth]{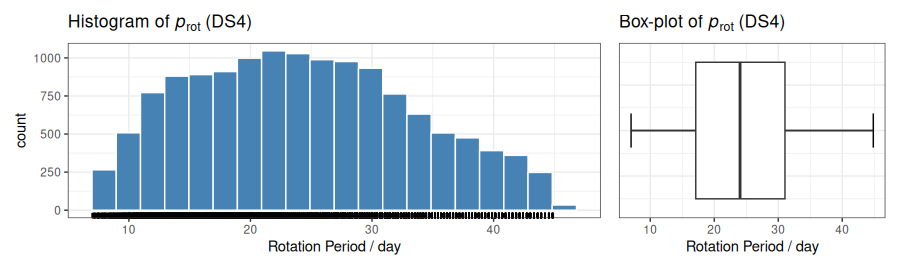}
  \end{subfigure}
  \vspace{5mm}
  \begin{subfigure}[b]{0.45\textwidth}
    \centering
    \includegraphics[width=\textwidth]{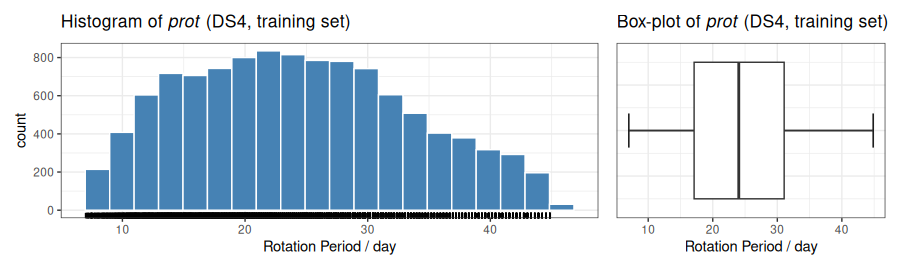}
  \end{subfigure}
  \caption{
    Similar to \cref{fig:prot-hist-bp-ds0}, but for the DS4 data set.
    The distributions are very similar to each other, approximately symmetrical, with means of \qty{24.44}{\day} and \qty{24.43}{\day}, and standard deviations of \qty{9.11}{\day} and \qty{9.07}{\day}, on DS4 and its training set, respectively.
  }
  \label{fig:prot-hist-bp-ds4}
\end{figure}
All variables present in DS0, including the response, $p_\text{rot}$ (the stellar rotation period, in days), are continuous.
The histograms and box-plots of the target variable, as extracted from DS0 (top panels) and its training set (bottom panels), are shown in \cref{fig:prot-hist-bp-ds0}.
The sample distributions are very similar to each other.
The aforementioned plots suggest unimodal, slightly platykurtic, right-skewed distributions, with several outliers towards large quantiles, corresponding to the largest stellar rotation periods, above approximately \qty{50}{\day} up to about \qty{100}{\day}.
The medians are equal to \qty{23.87}{\day}, with minima and maxima of \qtylist{0.53;98.83}{\day}, respectively.
The histograms show that the majority of the stars in the sample have rotation periods roughly between zero and \qty{50}{\day}, but the right tail of the distributions extends well beyond this interval.
Therefore, there is no statistical evidence of uniformity in the distribution of $p_\text{rot}$.

\Cref{fig:prot-hist-bp-ds4} is similar to \cref{fig:prot-hist-bp-ds0}, but for DS4.
The distributions on the full and training sets are very similar to each other.
There is no statistical evidence of uniformity in either set, although the distributions are flatter than their DS0 counterparts.
The plots suggest unimodal, platykurtic, slightly right-skewed distributions. 
For both DS4 and its training set, the median is \qty{24.03}{\day} and the minimum and maximum are \qtylist{7.00;44.85}{\day}, respectively;
the mean is equal to \qty{24.42}, and the standard deviations are \qtylist{9.10;9.12}{\day}, respectively.
From the histograms, we can see that the sample lacks stars with rotation periods of up to about \qty{12}{\day} and above about \qty{32}{\day}.
The fact that the distribution of $p_\text{rot}$ is not uniform affects the learning of the models, especially in the less represented regions.

\begin{figure*}[ht]
  \centering
  \includegraphics[width=\textwidth]{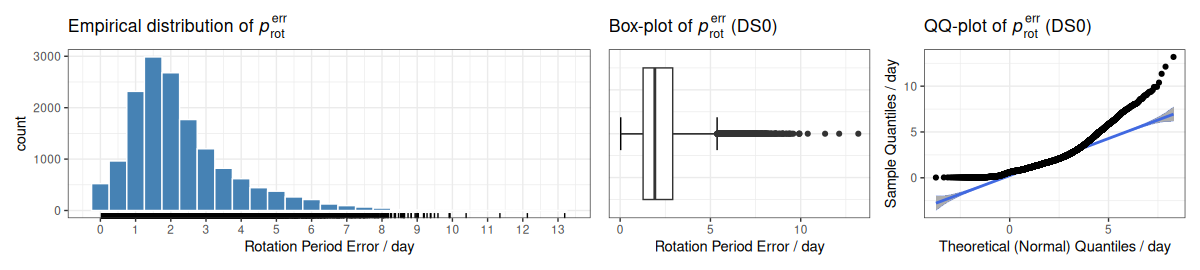}
  \caption{
    Histogram (left panel), box-plot (middle panel), and QQ-plot (right panel) of the rotation period errors, $p_\text{rot}^\text{err}$, on DS0.
    The distribution is unimodal, right-skewed, and slightly platykurtic, with a median of \qty{1.92}{\day}, ranging between \qtylist{0.03;13.19}{\day}.
  }
  \label{fig:prot-err-hist-bp-qq}
\end{figure*}

In DS0, after removing targets with rotation errors \qty{20}{\percent} larger than the corresponding rotation periods, as described in \cref{sec:data-engineering}, the distribution of the rotation period errors ($p_\text{rot}^\text{err}$, \cref{fig:prot-err-hist-bp-qq}) is unimodal, right skewed, and slightly platykurtic.
The errors vary between \qtylist{0.03; 13.19}{\day}, with a median of \qty{1.92}{\day}.
The outliers indicated by the box-plot are not significant, as all the errors are less than \qty{20}{\percent} of their corresponding rotation periods.
A significant number of points in the QQ-plot are outside the \qty{95}{\percent} confidence band, and the graph has two pronounced tails.
Therefore, there is no statistical evidence of normality in the distribution of $p_\text{rot}^\text{err}$.

\begin{figure}
  \centering
  \includegraphics[width=0.45\textwidth]{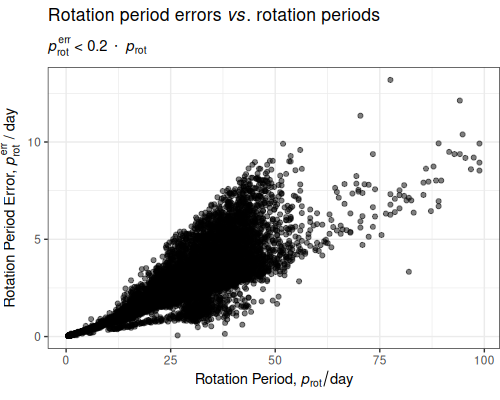}
  \caption{
    Rotation period errors, $p_\text{rot}^\text{err}$, as a function of the rotation periods, $p_\text{rot}$.
    The dispersion of the errors increases with the rotation periods.
  }
  \label{fig:prot-err-vs-prot}
\end{figure}

The scatter plot of the rotation period errors as a function of the rotation periods is illustrated in \cref{fig:prot-err-vs-prot}.
The majority of the measurements correspond to targets with rotation periods of less than 50 days.
Two branches can be identified, one starting at the lowest rotation periods and another starting at $p_\text{rot} \simeq \qty{10}{\day}$.
The nature of these branches is not known.
Both branches show a linear relationship between $p_\text{rot}^\text{err}$ and $p_\text{rot}$ without much dispersion up to a certain rotation period of about \qty{10}{\day} on the upper branch and \qty{27}{\day} on the lower branch.
Beyond these limits, the dispersion increases with $p_\text{rot}$.
The upper branch goes up to the maximum values of $p_\text{rot}$, while the lower branch stops at about \qty{55}{\day}.
An outlier at $p_\text{rot} = \qty{82}{\day}$ seems to belong to the lower branch, but as with the branches, its nature is unknown.
For rotation periods of around 50 days, the amplitude of the dispersion in the upper branch is about 7 days.
Increasing dispersion with $p_\text{rot}$ can lead to a decrease in model performance and be a source of outliers in the predictions.

\begin{figure}[ht]
  \centering
  \includegraphics[width=0.45\textwidth]{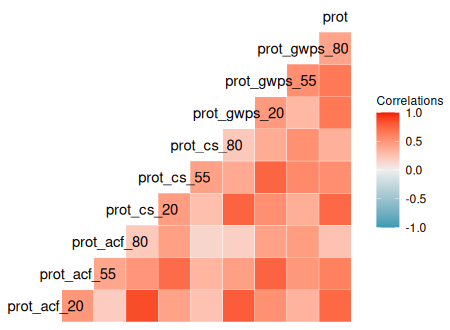}
  \caption{Correlations between the Prot variables and the response. In this case, all predictors are highly correlated with the target variable.}
  \label{fig:corr-prots}
\end{figure}

\begin{figure}[ht]
  \centering
  \includegraphics[width=0.45\textwidth]{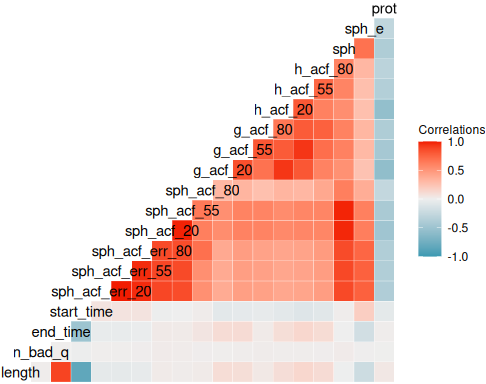}
  \caption{Correlations between the TS variables and the response. }
  \label{fig:corr-ts}
\end{figure}

\begin{figure}[ht]
  \centering
  \includegraphics[width=0.45\textwidth]{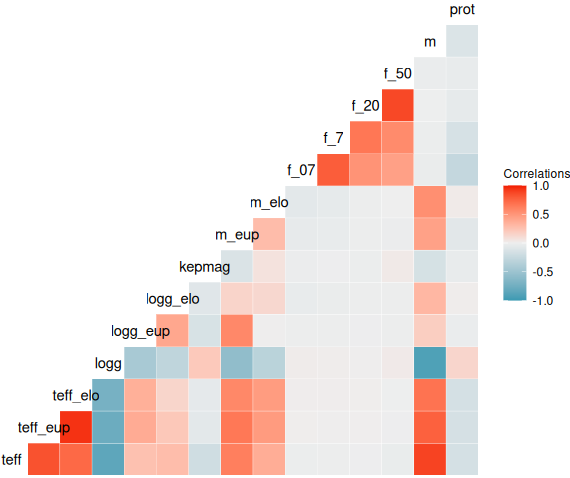}
  \caption{Correlations between the Astro variables and the response. }
  \label{fig:corr-astro}
\end{figure}

The correlations between the Prot, TS, and Astro families of variables and the response are shown in \cref{fig:corr-prots,fig:corr-ts,fig:corr-astro}.
The correlation between some of the variables belonging to the CS and GWPS families and the Astro group and Prot is shown in \cref{fig:corr-astro-cs,fig:corr-astro-gwps}.
Reddish colours indicate a positive correlation between any two features, blueish colours indicate a negative correlation, and white indicates no correlation at all.
As expected, most of the Prot variables are strongly positively correlated with the response, since the latter is obtained directly from the estimation of the former.
This was the motivation for creating the DS1 data set.
Some of the TS features are negatively correlated with the response (\cref{fig:corr-ts}), such as \texttt{h\_acf\_20}, \texttt{g\_acf\_20}, or \texttt{sph\_acf\_20}.
These predictors are expected to be important in building the models, as opposed to variables that have little correlation with $p_\text{rot}$, such as \texttt{start\_time}, \texttt{end\_time}, and \texttt{n\_bad\_q}.
All TS variables extracted from the ACF are highly correlated with each other, which in principle is not a problem because we used tree-based ensembles (see \cref{sec:problem-formulation}).
Classical Astro variables are not correlated with the response (\cref{fig:corr-astro}), apart from the mass, some cut-off frequencies of the Flicker in Power, and the effective temperature.
However, some features of the CS and GWPS families, such as \texttt{cs\_gauss\_}, \texttt{gwps\_gauss\_}, and their errors, are strongly correlated with the FliPer metrics \citep[see \cref{fig:corr-astro-cs,fig:corr-astro-gwps}]{bugnet2018fliper}.
This means that, in principle, any one of these variables could be replaced for the other without affecting how well the model performs.
For the CS and GWPS families of variables, we identified three groups of correlation with the response: low correlation, strong positive and strong negative correlation.
Examples of strong positive correlations are \texttt{cs\_gauss\_2\_1\_20}, \texttt{cs\_gauss\_3\_1\_20}, and \texttt{gwps\_gauss\_2\_1\_55}, and examples of strong negative correlations are \texttt{h\_cs\_20} and \texttt{sph\_gwps\_20}, just to name a few.
%%%%%%%%%%%%%%%%%%%%%%%%%%%%%%%%%%%%%%%%%%%%%%%%%%%%%%%%%%%%%%%%%%%%%%%%%%%%%%%%

\section{Methods}
\label{sec:methods}
\subsection{Problem Formulation}
\label{sec:problem-formulation}

Stellar rotation periods are positive real numbers, that can range from almost zero up to tens or hundreds of days \citep[\eg][]{santos2019surface,mcquillan2014rotation}, and so we framed the problem as a regression task.
Assuming that there is a relationship between a quantitative response $y$ and a set of $p$ predictors $\mathbf{x} = (x_1,\,x_2,\,\ldots,\,x_p)$, our problem can be described mathematically by the equation
\begin{equation}
  \label{eq:regression}
  y = f(\mathbf{x}) + e,
\end{equation}
where $f$ is an unknown fixed function of $\mathbf{x}$, and $e$ is a random \textit{error term} independent of $\mathbf{x}$, with zero mean.
The function $f$ can be estimated, but it is never fully known.
If the inputs $\mathbf{x}$ are readily available, given an estimate for $f$, $\hat f$, the output can be predicted by
\begin{equation}
  \label{eq:prediction}
  \hat y = \hat f(\mathbf{x}).
\end{equation}
In the context of this project, the target variable $y$ and the $p$ predictors are the rotation period, $p_\text{rot}$, and the set of variables described in \cref{sec:materials,tab:app-variables}, respectively, so that \cref{eq:regression} becomes
\begin{equation}
  \label{eq:main-eq}
  p_\text{rot} = f(x_1,\, x_2,\, \ldots,\, x_p) + e.
\end{equation}
In this equation, the index $p$ is an integer that varies up to 170, according to the number of variables that make up the data set used to train the model (DS0 to \replaced{DS7}{DS6}).
The models, trained by ML methods, can be expressed generically as
\begin{equation}
  \label{eq:model-eq}
  \hat{p}_\text{rot} = \hat f(x_1,\, x_2,\, \ldots,\, x_p),
\end{equation}
where $\hat f$ is the estimate of $f$.
We were not concerned with the exact form of $\hat f$, but rather with the accuracy of the predictions made by the models.

An important condition for us was that the models should simultaneously be robust, have good predictive performance, and be computationally cheap.
Given the tabular nature of the available data, we decided to use a tree-based ensemble method, specifically XGBoost, to tackle the research problem, as it is fast and has recently been shown to be the best for predictive tasks from structured data \citep{geron2017hands,raschka2017python}.

The proposed method aims to improve the predictive performance of existing models, such as the RF classifier published by \cite{breton2021rooster}, so that rotation periods can be estimated for thousands of K and M stars from the \textit{Kepler} catalogue, in an efficient and timely manner, with little human interaction. 
%%%-----------------------------------------------------------------------------

\subsection{The Extreme Gradient Boosting Approach}
\label{sec:approach-xgb}
Boosting can combine several weak learners, \ie models that predict marginally better than random, to produce a strong model, an ensemble with a superior generalised error rate.
\textit{Gradient Boosting} \citep[GB,][]{friedman2001greedy} is a method that can be applied to classification and regression tasks, based on the idea of \textit{steepest-descent minimisation}.
Given a loss function $L$ and a weak learner (\eg regression trees), it builds an additive model
\begin{equation}
  \label{eq:ada-additive}
  H(x_i) = \sum_k w_k h_k(x_i)\,,
\end{equation}
which attempts to minimise $L$.
In this equation, $w_k$ is the $k^\text{th}$ weight of the weak base model $h_k(x_i)$, and $x_i$ is the predictor $i$, as defined in \cref{sec:problem-formulation}.
The algorithm is typically initialised with the best estimate of the response, such as its mean in the case of regression, and it tries to optimise the learning process by adding new weak learners that focus on the residuals of the current ensemble.
The model is trained on a set consisting of the cases $(x_i, r_i)$, where $r_i$ is the $i^\text{th}$ residual, given by
\begin{equation}
  \label{eq:gb}
  r_i = -\frac{\partial L(y_i, f(x_i))}{\partial f(x_i)}\,.
\end{equation}
The residuals thus provide the gradients, which are easy to calculate for the most common loss functions.
The current model is added to the previous one, and the process continues until a stopping-condition is met (\eg the number of iterations, specified by the user).

We used XGBoost, a successful implementation of GB, which was designed with speed, efficiency of computing resources, and model performance in mind.
The algorithm is suitable for structured data.
It is able to automatically handle missing values, it supports parallelisation of the tree-building process, it automatically ensures early stopping if the training performance does not evolve after some predetermined iterations, and it allows the training of a model to be resumed by boosting an already fitted model on new data \citep{chen2016xgboost}.
The method implements the GB algorithm with minor improvements in the objective function.
A prediction of the response is obtained from a tree-ensemble model built from $K$ additive functions:
\begin{equation}
  \label{eq:xgb-model}
  \hat y_i = \sum_{k=1}^K f_k(x_i),
\end{equation}
where the function $f_k$ belongs to the functional space $\mathcal{F}$ of all possible decision trees.
Each $f_k$ represents an independent tree learner.
The final prediction for each example is given by the aggregation of the predictions of the individual trees.
The set of functions in the ensemble is learned by minimising a regularised objective function, $\mathcal{L}(y_i,\,\hat y_i)$, which consists of two parts:
\begin{equation}
  \label{eq:xgb-objective-function}
  \mathcal{L}(y_i,\,\hat y_i) = \sum_{i=1}^n\ell(y_i,\, \hat y_i) + \sum_{k=1}^K \omega(f_k).
\end{equation}
The first term on the right-hand side of \cref{eq:xgb-objective-function} is the \textit{training loss}, $L(y_i,\, \hat y_i)$, and the second term is the \textit{regularisation term}.
Here, $\ell$ is a differentiable convex loss function that measures the difference between the predicted and the true values of the response.
The training loss measures how good the model is at predicting using the training data\replaced{, and it may be selected from a number of performance metrics, some of which will be described in detail in \cref{sec:performance}.}{A common choice for $L$ is the \textit{mean squared error.}}
The regularisation term controls (penalises) the complexity of the model.
It helps to smooth the final learning weights, to avoid overfitting.
In practice, the regularised objective function tends to select models that use simple and predictive functions \citep{chen2016xgboost}.
A model is learned optimising the objective function above.

The XGBoost algorithm has several \textit{hyperparameters}, that are not considered in \cref{eq:model-eq}.
Hyperparameters are tuning parameters that cannot be estimated directly from the data, and that are used to improve the performance of an ML model \citep{kuhn2013applied}.
They are parameters of a learning algorithm, not of a model, and are mostly used to control how much of the data is fitted by the model, so that, ideally, the true structure of the data is captured by the model, but not the noise.
Optimising them using a resampling technique is crucial and the best way to build a robust model with good predictive performance.
They are set in advance and remain constant throughout the training process.
Hyperparameters are not affected by the learning algorithm, but they do affect the speed of the training process \citep{geron2017hands,raschka2017python}.
There is no formula for calculating the optimal value nor a unique rule for tuning the parameters used to estimate a given model.
The optimal configuration depends on the data set, and the best way to build a model is to test different sets of hyperparameter values using resampling techniques.
\Cref{sec:experimental-design} describes the strategy we used to optimise the XGBoost hyperparameters.
%%%-----------------------------------------------------------------------------

\subsection{Performance Assessment}
\label{sec:performance}
The performance of the ML models built with the data sets described in \cref{sec:data-engineering} was assessed by quantifying the extent to which the predictions were close to the true values of the response for the set of observations.
Given the regression nature of the problem in this project, we used six metrics to assess the predictive quality and the goodness of fit of the learners:
\begin{enumerate*}[label=(\alph*), font=\itshape]
  \item the \textit{root mean squared error} (RMSE) and the \textit{mean absolute error} (MAE);
  \item an interval-based error or residual, $\epsilon_{\Delta, \delta}$;
  \item an interval-based ``accuracy'', $\text{acc}_\Delta$, calculated using the interval-based error;
  \item the mean of the absolute values of the residuals, $\mu_\text{err}$; and
  \item the adjusted coefficient of determination, $R^2_\text{adj}$.
\end{enumerate*}
\added{Furthermore, the error bars associated with the predictions were estimated in accordance with the specifications outlined in \cref{sec:error-bars}.}
The RMSE, the MAE, and $\epsilon_{\Delta, \delta}$ were used to measure the statistical dispersion, while $\text{acc}_\Delta$ and $R^2_\text{adj}$ were used to assess the predictive performance of the models.
The mean of the residuals, $\mu_\text{err}$, was used simultaneously as a measure of the dispersion and of the quality of the models, in the sense that we could infer from it how wrong the models were on average.

During the learning phase, a model's performance was estimated using the $k$-fold cross-validation (CV), where a subset of observations was used to fit the model, and the remaining instances were used to assess its quality.
This process was repeated several times, and the results were aggregated and summarised.

After the training phase, predictions were made on each of the testing sets, and the resulting rotation periods were compared with the reference values contained in the S19 catalogue.
The mean of the relative absolute values of the residuals was used as a measure of goodness of fit, giving us an idea of how wrong the models were on average.
The RMSE and the MAE were used together to measure the differences between the predicted and reference values of stellar rotation periods.
They were particularly useful in assessing the performance of the models on the training and testing sets, and thus to better tuning them in order to avoid overfitting---large differences in the RMSE and/or MAE between the training and testing sets are usually an indication of overfitting.
The interval-based accuracy allowed us to convert the assessment of a regression problem into the evaluation of a classification result.
This in turn allowed us to compare the predictive performance of our models with those trained by \cite{breton2021rooster}.
Since the latter assessed their results within a \qty{10}{\percent} interval of the reference values, we used the \qty{10}{\percent}-accuracy, acc$_{0.1}$, as a benchmark.
When calculating the interval accuracies, it should be noted that the width of the intervals centred on the reference values increases with the rotation period, while the width of the intervals around the corresponding rotation frequencies does not vary significantly with the frequency, except in the low-frequency range.
Since the distribution of the amplitudes of the frequency intervals is more uniform than that of the amplitudes of the period intervals, the rotation periods were first converted to rotation frequencies in order to reduce the effect of the increasing width of the accuracy intervals with the period, which would affect the comparisons and the assessment of the quality of the models.

The following is a brief description of each of the above metrics and the $k$-fold CV technique.

\paragraph{Root Mean Squared Error.}
The most commonly used metric to assess the performance of a model in a regression setting is the \textit{mean squared error} (MSE).
This is defined as
\begin{equation}
  \label{eq:mse-ml}
  \text{MSE} = \frac{1}{n}\sum_{i=1}^{n}\left[ (y_i - \hat y_i)^2 \right],
\end{equation}
where $\hat y_i = \hat f(\mathbf{x}_i)$ is the prediction for the $i^\text{th}$ observation given by $\hat f$, and $n$ is the number of cases present in the data set.
The MSE is a measure of the distance between a point estimator $\hat y$ and the real value or ground truth $y$.
It is usually interpreted as how far from zero the \textit{errors} or \textit{residuals} are on average, \ie its value represents the average distance between the model predictions and the true values of the response \citep{kuhn2013applied}.
In general, a smaller MSE indicates a better estimator \citep{ramachandran2020mathematical}: the MSE will be small when the predictions and the responses are slightly different, and will typically be large when they are not close for some cases.
The MSE can be computed on both the training and testing data, but we are mostly interested in measuring the performance of the models on unseen data (the testing sets)---the model with the smallest testing MSE has the best performance.
If a learner produces a small training MSE but a large testing MSE, this is an indication that it is overfitting the data, and therefore a less flexible model would produce a smaller testing MSE \citep{james2013introduction}.
A common measure of the differences between estimates and actual values is the \textit{root mean squared error} (RMSE), which is no more than the square root of the MSE:
\begin{equation}
  \label{eq:rmse}
  \text{RMSE} = \sqrt{\text{MSE}}\,.
\end{equation}
The RMSE has the advantage over the MSE that it has the same units as the estimator.
For an unbiased estimator, it is equal to the standard deviation.
Similarly to the MSE, smaller values of the RMSE generally indicate a better estimator, but as this metric is dependent on the scale of the variables used, comparisons are only valid between models created with the same data set \citep{hyndman2006another}.

\paragraph*{Mean Absolute Error.}
The \textit{mean absolute error} (MAE) of an estimator $\hat y = f(\mathbf{x})$ is the average of the absolute values of the errors:
\begin{equation}
  \label{eq:mae-ml}
  \text{MAE} = \frac{1}{n} \sum_{i=1}^n \lvert y_i - \hat y_i \rvert\,,
\end{equation}
where all of the quantities have the same meaning as before.
The MAE is a measure of the error between the prediction and the ground truth.
It uses the same scale as the estimates and the true values.
Therefore, it cannot be used to make comparisons between models built on different data sets.
The MAE has the advantage over the RMSE that it is affected by the error in direct proportion to its absolute value \citep{pontius2008components}.
When used together, the RMSE and MAE can diagnose variations in the residuals in a set of predictions.
For each predictive model and sample under consideration,
\begin{equation}
  \label{eq:rmse-mae}
  \text{RMSE} \geqslant \text{MAE}\,.
\end{equation}
The variance of the individual errors increases with the difference between the RMSE and the MAE, and all the errors are of the same size when both metrics are equal to each other.

\paragraph*{Mean Absolute Value of the Relative Error.}
The \textit{mean absolute value of the relative error} (MARE) is given by
\begin{equation}
  \label{eq:mare}
  \text{MARE} = \mu_\text{err} = \frac{1}{n} \sum_{i=1}^n \left\lvert \frac{y_i - \hat y_i}{y_i} \right\rvert,
\end{equation}
where $n$ is the number of observations, $y_i = f(\mathbf{x}_i)$ and $\hat y_i = \hat f (\mathbf{x}_i)$.
It is similar to the MAE, but the mean absolute value is calculated on the relative residuals rather than the magnitude of the errors.
The MARE is a measure of the model's mean relative error, giving a percentage estimate of how wrong the model is on average.

\begin{figure}[!ht]
  \begin{center}
    \includegraphics[width=0.8\linewidth]{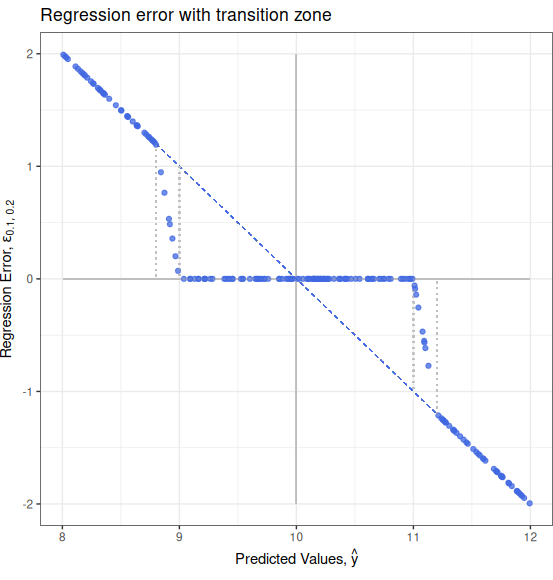}
    \caption{
      Graphical representation of the interval-based error function computed on simulated data, for a fixed ground truth of 10.
      The plot shows the \qty{10}{\percent}-width error function with a transition zone of \qty{20}{\percent} of the half width of the error-interval.
      The blue dashed line indicates the residuals and the grey vertical dotted lines indicate the transition zones.
    }
    \label{fig:regression-error}
  \end{center}
\end{figure}

\paragraph*{Interval Accuracy.}
We developed an interval-based error function and an ``accuracy'' metric to bridge the gap between regression and classification settings.
The \textit{interval-based error}, $\epsilon_{\Delta, \delta}$, estimates the prediction error in a regression task when we are willing to accept an error of $100\times\Delta\,\%$, and it is given by:
\begin{equation}
  \label{eq:error-int}
  \epsilon_{\Delta, \delta} =%
  \begin{cases}
    0, & \lvert \epsilon_i \rvert \leqslant \Delta \cdot y_i \\
    \left[ (1 + \Delta) \cdot y_i - \hat y_i \right]\frac{1 + \delta}{\delta}, & \Delta \cdot y_i < \epsilon_i \leqslant \Delta \cdot (1 + \delta) \, y_i \\
    \left[ (1 - \Delta) \cdot y_i - \hat y_i \right]\frac{1 + \delta}{\delta}, & - \Delta \cdot (1 + \delta)\, y_i \leqslant \epsilon_i < - \Delta y_i \\
    \epsilon_i, & \text{otherwise,}
  \end{cases}
\end{equation}
where $\epsilon_i = y_i - \hat y_i$, $y_i = f(\mathbf{x}_i)$ and $\hat y_i = \hat f(\mathbf{x}_i)$ are the $i^\text{th}$ ground truth and predicted values, respectively, $\Delta$ is the fractional width of the \textit{zeroing interval}, and $\delta$ is the fraction of $\Delta\cdot y$ that defines a transition zone, where the error varies linearly between zero and the typical residual.
When $\Delta$ is zero, the metric returns the simple residuals.
The second and third branches of \cref{eq:error-int} correspond to the transition zone.
We used a linear function to join the residuals to the error interval, but any other function could be used, such as a logistic function or a hyperbolic tangent, just to name a few.
\Cref{fig:regression-error} shows the interval-based error function calculated on simulated data, when an error of \qty{10}{\percent} is considered acceptable, \ie when $\Delta = 0.1$.
The transition zone is defined for $\delta = 0.2$, corresponding to \qty{20}{\percent} of $\Delta \cdot y$.
The predictions were obtained from a uniform distribution of \num{200} values varying between 0 and 2 around a fixed ground truth value of 10.
Similar to \cref{eq:mare}, this error function, when normalised to the reference values, can be used to approximate how much a model is wrong on average for a given \textit{error-interval} width.

An \textit{interval-based} ``accuracy'', $\text{acc}_\Delta$, can be obtained from the interval-based residuals of \cref{eq:error-int}:
\begin{equation}
  \label{eq:acc-int}
  \text{acc}_\Delta = \frac{1}{n} \sum_{i=1}^n I \left( \epsilon_{\Delta, \delta} = 0 \right),
\end{equation}
where $\Delta$, $y_i$, and $\hat y_i$ are as before, and $n$ is the number of observations.\footnote{%
  Note that while $\epsilon_{\Delta, \delta}$ depends on $\delta$, acc$_\Delta$ does not.
}
The indicator function is equal to 1 when the interval-based regression error is zero, \ie whenever $\lvert y_i - \hat y_i \rvert \leqslant \Delta\cdot y_i$, and zero otherwise;
that is, we consider an \textit{event} every time the error in \cref{eq:error-int} is equal to zero, and a \textit{non-event} otherwise.

\paragraph{Adjusted Coefficient of Determination.}
We used the \textit{adjusted coefficient of determination}, $R^2_\text{adj}$, as a fair measure of a model's goodness of fit, to characterise its predictive ability:
\begin{equation}
  \label{eq:r2adj}
  R^2_\text{adj} = 1 - \frac{(n - 1)}{n - p - 1} \cdot (1 - R^2),
\end{equation}
where $R^2$ is the coefficient of determination, $n$ is the number of observations, and $p$ is the number of predictor variables \citep{baron2019probability}.

\begin{figure}[!ht]
  \centering
  \includegraphics[width=0.95\linewidth]{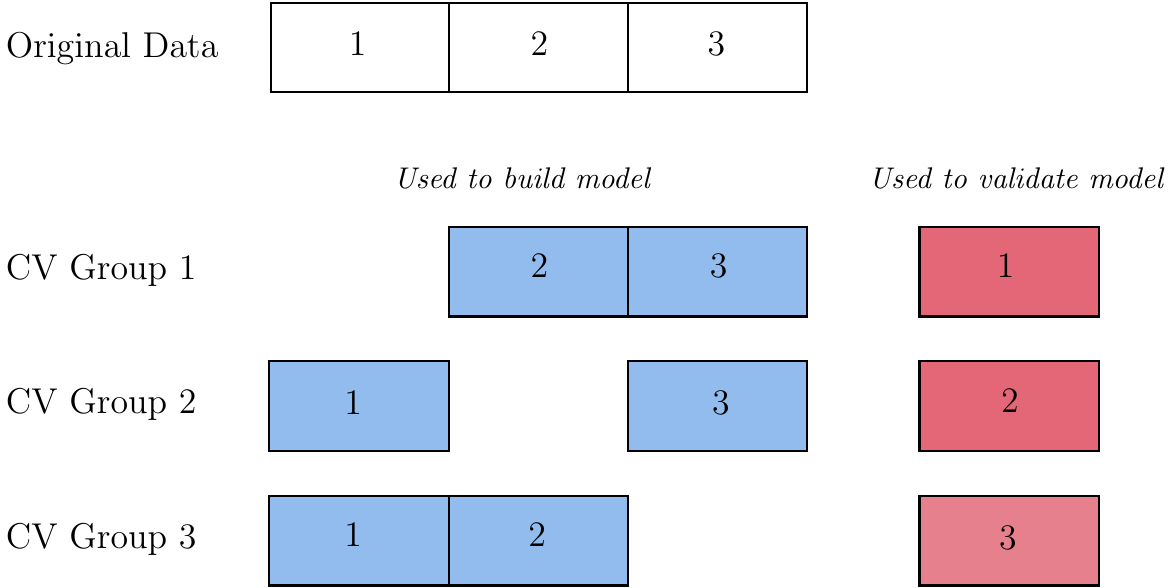}
  \vspace{2mm}
  \caption{
    A schematic of a 3-fold CV process.
    The original training data are divided into three non-overlapping subsets.
    Each fold is held out in turn while models are fitted to the remaining training folds.
    Performance metrics, such as MSE or $R^2$, are estimated using the retained fold in a given iteration.
    The CV estimate of the model performance is given by the average of the three performance estimates.
  }
  \label{fig:k3fcv}
\end{figure}

\mathversion{bold}
\paragraph*{$k$-fold cross-validation.}
\label{sec:kfcv}
\mathversion{normal}
We used the $k$-\textit{fold cross-validation} as a resampling technique during the training phase of the models, in which the training instances were randomly partitioned into $k$ non-overlapping sets or folds of approximately equal size.
One of the folds was kept as the validation set, and the remaining folds were combined into a training set to which a model was fitted.
After assessing the performance of the model on the validation fold, the latter was returned to the training set, and the process was repeated, with the second fold as the new validation set, and so on and so forth \citep{kuhn2013applied,hastie2009elements}.
The $k$ performance estimates were summarised with the mean and the standard error, and they were used to tune the model parameters.
The testing error was estimated by averaging out the $k$ resulting MSE estimates.
The bias of the technique, \ie the difference between the predictions and the true values in the validation set, decreases with $k$.
Typical choices for $k$ are 5 and 10, but there is no canonical rule.
\Cref{fig:k3fcv} illustrates an example of a CV process with $k = 3$.

\added{
  \subsubsection{Estimation of Error Bars}
  \label{sec:error-bars}
  In order to estimate the errors associated with the predictions made on the testing set, we employed a straightforward methodology, whereby the predictions were binned into uniform unit intervals centred on integer values spanning between the minimum and the maximum.
  The standard deviation within each bin was then taken as the error bar.
}
%%%%%%%%%%%%%%%%%%%%%%%%%%%%%%%%%%%%%%%%%%%%%%%%%%%%%%%%%%%%%%%%%%%%%%%%%%%%%%%%

\section{Experimental Design}
\label{sec:experimental-design}
We applied XGBoost to build regressor models, each of which was trained using the hold-out method, where the data sets described in \cref{sec:data-engineering} were split into training and testing subsets, in an 80-to-20 percent ratio.
\added{%
  This proportion was reflected in the following distribution: in the case of DS0, DS1, DS2, and DS7, the training and testing sets comprised \num{11930} and \num{2983} instances, respectively;
  DS3 and DS5 consisted of \num{11484} observations in the training set and \num{2872} elements in the testing set;
  finally, for DS4 and DS6, \num{10881} instances were for the training set and \num{2721} for the testing one.
}%
We converted all training and testing sets to \textit{XGB dense matrices} for efficiency and speed, as recommended by the authors of the \textbf{xgboost} package \citep{chen2016xgboost}.
During the training phase, we used a 10-fold CV with five repetitions, where the best values for the hyperparameters were sought using a search grid on the training set.
In all cases, the testing sets remained unseen by the models until the prediction and model evaluation phase.
That is, the instances belonging to the testing sets were at no time part of the model learning process, behaving as new, unknown observations, as it would happen in a real-life scenario.

The experiment was carried out in two iterations.
In the first iteration, we searched for the best values of the hyperparameters using a two-step grid search:
firstly, the best parameters were searched for within a set of possible values;
the search was then refined on the learning rate and the subsample ratio of training instances, by varying the parameter in a finer grid centred on the best value found in the previous step.
\replaced{Typically, n}{N}o more than \replaced{1500}{1000} iterations were run for each submodel, and we activated a stopping criterion, where the learning for a given model would stop if the performance did not improve for five rounds.
The grid values for each parameter were as follows:
\begin{itemize}
  \item \texttt{colsample\_bytree}: the fraction of columns to be randomly subsampled when constructing each tree; it was searched within the set $\{1/3,\, 1/2,\, 2/3\}$\deleted{, also during the first grid search}.\footnote{
    Two other hyperparameters belonging to this family of subsampling columns were initially optimised: \texttt{colsample\_bylevel} and \texttt{colsample\_bynode}.
    However, no improvement in model performance was observed when they were included, and as they work cumulatively, only \texttt{colsample\_bytree} was retained in the optimisation process in the end.
  }
  \item \texttt{eta}, \added{$\eta$}: the \textit{learning rate}, was first searched within the values $\{0.03,\, 0.04,\, 0.05,\, 0.06,\, 0.07,\, 0.08,\, 0.09,\, 0.1\}$; the search was then refined by looking for five values centred on the best $\eta$ found in the first iteration, each \num{0.01} apart.
  \item \texttt{gamma}\added{, $\gamma$}: the minimum loss reduction required to further partition a leaf node of the tree; it was chosen within the set of values $\{\added{0}, 1,\, 3,\, 5,\, 10\}$.
  \item \texttt{max\_depth}: the maximum depth per tree; during the first iteration, the best value was chosen from the set of integers $\{5\mathsf{L},\, 6\mathsf{L},\, 7\mathsf{L}\}$\added{; in the subsequent iteration, the optimal value was selected from the set $\{7\mathsf{L},\, 8\mathsf{L},\, 9\mathsf{L}\}$}.
  \item \texttt{min\_child\_weight}: the minimum sum of instance weights required in a child node, which in a regression task is simply the minimum number of instances needed in each node; this parameter was searched for within the grid of integers $\{1\mathsf{L},\, 5\mathsf{L},\, 10\mathsf{L},\, 20\mathsf{L}\}$.
  \item \texttt{subsample}: the subsample ratio of the training instances; we opted for both stochastic and regular boosting, and so during the first grid search, \texttt{subsample} was chosen within eleven possible values, between \num{0.5} and \num{1}, varying in steps of \num{0.05}; the best value of the parameter found in the first grid search was then refined by searching for the \replaced{optimal}{best} of five values centred on this best value or, if the latter was 1, by searching for the best of three values less than 1, each \num{0.05} apart in all cases.
\end{itemize}
We tried other XGB hyperparameters, but we found that these had the greatest impact on the performance of the model.
The grid produced a maximum of \num{12672} + 25 submodels during training, not counting the 10-fold CV with five repetitions.

In the second iteration, we refined the grid search, by using three values per hyperparameter, centred on the optimum for each parameter found in the first iteration.
The grid is given in \cref{tab:grid-iter2}.
The rationale for iteration 2 was to search for two additional values for each hyperparameter around the optimal value found in iteration 1.
In the case of \texttt{eta} and \texttt{subsample}, care was taken not to exceed the allowed limits for these hyperparameters.
The training produced a maximum of 729 + 20 submodels, not counting the CV.
\begin{table*}[ht]
  \caption{
    Set of possible values used for the hyperparameter grid search in iteration 2.
  }
  \label{tab:grid-iter2}
  \centering
  \resizebox{\linewidth}{!}{
    \begin{tabular}{lcccccccc}
      \toprule
      Hyperparameter & DS0 & DS1 & DS2 & DS3 & DS4 & DS5 & DS6 & DS7 \\
      \midrule
      \texttt{colsample\_bytree} & 3/5, 2/3, 3/4 & \repdel{2/5}{3/5}, \repdel{1/2}{2/3}, \repdel{3/5}{3/4} & \repdel{1/5}{3/5}, \repdel{1/3}{2/3}, \repdel{2/5}{3/4} & 2/5, 1/2, 3/5 & 2/5, 1/2, 3/5 & 2/5, 1/2, 3/5 & 2/5, 1/2, 3/5 & 2/5, 1/2, 3/5 \\
      \addlinespace[1mm]
      \texttt{eta} & \begin{tabular}{@{}c@{}}0.03\\0.04\\0.05\end{tabular} & \begin{tabular}{@{}c@{}}0.01\\0.02\\0.03\end{tabular} & \begin{tabular}{@{}c@{}}\repdel{0.01}{0.02}\\\repdel{0.02}{0.03}\\\repdel{0.03}{0.04}\end{tabular} & \begin{tabular}{@{}c@{}}\repdel{0.02}{0.04}\\\repdel{0.03}{0.05}\\\repdel{0.04}{0.06}\end{tabular} & \begin{tabular}{@{}c@{}}0.02\\0.03\\0.04\end{tabular} & \begin{tabular}{@{}c@{}}\repdel{0.02}{0.01}\\\repdel{0.03}{0.02}\\\repdel{0.04}{0.03}\end{tabular} & \begin{tabular}{@{}c@{}}\repdel{0.01}{0.02}\\\repdel{0.02}{0.03}\\\repdel{0.03}{0.04}\end{tabular} & \begin{tabular}{@{}c@{}}0.01\\0.02\\0.03\end{tabular} \\
      \addlinespace[1mm]
      \texttt{gamma} & 0, 1, 3 & 5, 10, 15 & 1, 3, 5 & \repdel{0}{5}, \repdel{1}{10}, \repdel{3}{15} & \repdel{0}{1}, \repdel{1}{3}, \repdel{3}{5} & \repdel{0}{5}, \repdel{1}{10}, \repdel{3}{15} & 0, 1, 3 & \repdel{5}{1}, \repdel{10}{3}, \repdel{15}{5} \\
      \addlinespace[1mm]
      \texttt{max\_depth} & 7L, 8L, 9L & 7L, 8L, 9L & 7L, 8L, 9L & 7L, 8L, 9L & 7L, 8L, 9L & 7L, 8L, 9L & 7L, 8L, 9L & 7L, 8L, 9L \\
      \addlinespace[1mm]
      \texttt{min\_child\_weight} & 15L, 20L, 25L & \repdel{15}{5}L, \repdel{20}{10}L, \repdel{25}{15}L & \repdel{15}{5}L, \repdel{20}{10}L, \repdel{25}{15}L & \repdel{15}{5}L, \repdel{20}{10}L, \repdel{25}{15}L & \repdel{5}{1}L, \repdel{10}{5}L, \repdel{15}{10}L & 5L, 10L, 15L & 5L, 10L, 15L & \repdel{5}{1}L, \repdel{10}{5}L, \repdel{15}{10}L \\
      \addlinespace[1mm]
      \texttt{subsample} & 0.90, 0.95, 1.0 & 0.90, 0.95, 1.0 & 0.90, 0.95, 1.0 & 0.90, 0.95, 1.0 & \repdel{0.85}{0.90}, \repdel{0.90}{0.95}, \repdel{0.95}{1.0} & \repdel{0.85}{0.90}, \repdel{0.90}{0.95}, \repdel{0.95}{1.0} & 0.90, 0.95, 1.0 & \repdel{0.85}{0.90}, \repdel{0.90}{0.95}, \repdel{0.95}{1.0} \\
      \bottomrule
    \end{tabular}
  }
\end{table*}

After being trained by performing a grid search on the selected hyperparameters, the models were evaluated, and the importance of each feature was assessed in terms of the predictive power of the model, using the variance of the responses.
The following sections present results and analysis for all models obtained during both iterations.
%%%%%%%%%%%%%%%%%%%%%%%%%%%%%%%%%%%%%%%%%%%%%%%%%%%%%%%%%%%%%%%%%%%%%%%%%%%%%%%%

\section{Results}
\label{sec:results}
In \cref{tab:xgb-best-params}, we present the results of the parameter optimisation carried out in iterations 1 and 2.
As far as the hyperparameters are concerned, we highlight the following points by comparing the results of \cref{tab:xgb-best-params} with the original grid of \cref{sec:experimental-design}:
\begin{itemize}
  \item \texttt{colsample\_bytree}
    \begin{itemize}
      \item[\tiny\textbullet] In iteration 1, the optimal value was\deleted{$2/3$ for the models built with the largest data sets (DS0 to DS2) and} $1/2$ \added{or less} for \added{all} the \deleted{remaining} models \replaced{except DS0}{(DS3 to DS7)}; thus, \replaced{in the data sets comprising less predictors}{with the exception of DS2}, fewer variables were randomly subsampled \deleted{as the number of predictors decreased};
      \item[\tiny\textbullet] In iteration 2, the tendency to reduce the number of randomly sampled variables with smaller sets remained.
    \end{itemize}
  \item \texttt{eta}
    \begin{itemize}
      \item[\tiny\textbullet] In the first iteration, \replaced{$\eta$}{it} remained below 0.05, with \replaced{0.02}{0.03} being the most common value; so, smaller values were prioritised; this came at the cost of longer convergence times and the risk of overfitting, so we relied on \texttt{gamma} for a more conservative approach;
      \item[\tiny\textbullet] In iteration 2, the most common learning rate value was 0.02, and \added{only} in \replaced{one}{three} situation\deleted{s} \added{was} the best value \replaced{0.03}{was 0.01}. 
    \end{itemize}
  \item \texttt{gamma}
    \begin{itemize}
      \item[\tiny\textbullet] In iteration 1, \replaced{$\gamma$}{it} was equal to \added{5 or} 10 in three models, imposing the most conservative constraint of the options offered --- this happened for\deleted{both} the\deleted{highest (\num{0.05}) and} lowest (\replaced{0.02}{0.01}) optimal value\deleted{s} of the learning rate; \added{nevertheless, this lowest $\eta$ value also correlated with the low $\gamma$ value of 1 in four models; this suggests, as expected, that the optimal values for these hyperparameters are also dependent on the data set in question and the values of the other hyperparameters;}
      \item[\tiny\textbullet] In the second iteration, \replaced{$\gamma$ was observed to vary between 0 and 5, with 3 being the most common value;}{it was either equal to 1 or 5, since the constraints imposed on the models were typically less conservative than \added{those} in the previous case.} \added{in DS5, no constraints were applied to the model, and in no case was the most conservative value of 10 imposed.}
    \end{itemize}
  \item \texttt{max\_depth} 
    \begin{itemize}
      \item[\tiny\textbullet] In iteration 1, it was always equal to 7, which is the maximum value offered by the grid;
      \item[\tiny\textbullet] Iteration 2 offered cases with higher values (8 and 9) and, except for \added{the} DS1\replaced{,}{and} DS2\added{, and DS4} models, 9 was the best value\deleted{; it was never equal to 7}.
    \end{itemize}
  \item \texttt{min\_child\_weight} 
    \begin{itemize}
      \item[\tiny\textbullet] In the first iteration, the most common value was \replaced{20}{10} and the minimum value was \replaced{10}{5}; therefore, highly regularised models were generated, with smaller trees than when this hyperparameter was equal to 1 (the default value) --- this limited a possible perfect fit for some observations, but made the models less prone to overfitting \added{and compensated for smaller values of $\eta$ and $\gamma$};
      \item[\tiny\textbullet] In the second iteration,\deleted{similarly to the previous case,} \replaced{DS1, DS2, and DS3 were}{DS0 was} the model\added{s} with the highest degree of regularisation; the most common value\added{s} \replaced{were 15 and 20}{remained 10}, and with \replaced{them}{it} the trend towards models with smaller trees.
    \end{itemize}
  \item \texttt{subsample} 
    \begin{itemize}
      \item[\phantom{-}] It was equal or close to 1.0 \added{(never below 0.90)} in both iterations, and so regular boosting was prioritised, despite the range of values suitable for stochastic boosting available in the grid.
    \end{itemize}
\end{itemize}
\begin{table*}[ht]
  \caption{
    Best set of values for the XGB hyperparameters after performing a grid search.
    The \texttt{nrounds} parameter indicates the number of iterations that the XGB algorithm should perform to obtain the best model using the optimal values of the hyperparameters for the given data set.
  }
  \label{tab:xgb-best-params}
  \centering
  \resizebox{\linewidth}{!}{
    \begin{tabular}{lccccccccccccccccc}
      \toprule
       & \multicolumn{8}{c}{Iteration 1} & & \multicolumn{8}{c}{Iteration 2} \\
      \cmidrule{2-9}\cmidrule{11-18}
      Hyperparameter & DS0 & DS1 & DS2 & DS3 & DS4 & DS5 & DS6 & DS7 && DS0 & DS1 & DS2 & DS3 & DS4 & DS5 & DS6 & DS7\\
      \midrule
      %  & & & & & & & && & & & & & & \\
      \texttt{colsample\_bytree} & 2/3 & \repdel{1/2}{2/3} & \repdel{1/3}{2/3} & 1/2 & 1/2 & 1/2 & 1/2 & 1/2 && 3/4 & \repdel{2/5}{3/4} & \repdel{2/5}{2/3} & \repdel{1/2}{3/5} & 1/2 & \repdel{2/5}{1/2} & \repdel{2/5}{1/2} & 1/2 \\
      \addlinespace[1mm]
      \texttt{eta} & 0.04 & 0.02 & \repdel{0.02}{0.03} & \repdel{0.03}{0.05} & \repdel{0.02}{0.03} & \repdel{0.02}{0.01} & \repdel{0.02}{0.03} & 0.02 && 0.03 & 0.02 & \repdel{0.02}{0.01} & 0.02 & 0.02 & \repdel{0.02}{0.01} & 0.02 & \repdel{0.02}{0.01} \\
      \addlinespace[1mm]
      \texttt{gamma} & 1 & 10 & 3 & \repdel{1}{10} & \repdel{5}{3} & \repdel{5}{10} & 1 & \repdel{1}{3} && 1 & 5 & 3 & \repdel{3}{5} & \repdel{3}{5} & \repdel{0}{5} & 1 & 5 \\
      \addlinespace[1mm]
      \texttt{max\_depth} & 7 & 7 & 7 & 7 & 7 & 7 & 7 & 7 && 9 & \repdel{7}{8} & \repdel{8}{9} & 9 & \repdel{8}{9} & 9 & 9 & 9 \\
      \addlinespace[1mm]
      \texttt{min\_child\_weight} & 20 & \repdel{20}{10} & \repdel{20}{10} & \repdel{20}{10} & \repdel{20}{5} & \repdel{20}{10} & 10 & \repdel{10}{5} && 15 & \repdel{25}{10} & \repdel{25}{15} & \repdel{25}{10} & \repdel{15}{10} & \repdel{15}{10} & 10 & 10 \\
      \addlinespace[1mm]
      \texttt{subsample} & 1.0 & 1.0 & \repdel{0.90}{1.0} & \repdel{0.95}{1.0} & \repdel{0.90}{1.0} & \repdel{0.90}{0.95} & 0.95 & 0.95 && 1.0 & 1.0 & \repdel{1.0}{0.90} & 1.0 & 0.90 & \repdel{0.90}{1.0} & 0.90 & \repdel{0.90}{0.95} \\
      %  & & & & & & & && & & & & & & \\
      \addlinespace[3.5mm]
      \texttt{nrounds} & 359 & \repdel{1422}{753} & \repdel{1453}{492} & \repdel{877}{185} & \repdel{1249}{362} & \repdel{881}{926} & \repdel{1275}{489} & \repdel{1153}{769} && 470 & \repdel{1373}{657} & \repdel{1516}{1410} & \repdel{1285}{499} & \repdel{958}{451} & \repdel{1078}{924} & \repdel{1114}{806} & \repdel{1089}{998} \\
      \bottomrule
    \end{tabular}
  }
\end{table*}

The models were built using the optimal sets of the hyperparameters found by the grid search, and the algorithm was stopped after \texttt{nrounds} iterations, as given in the last row of \cref{tab:xgb-best-params}.
We had set the maximum number of possible iterations during the learning phase to \replaced{1500.}{1000,} \deleted{but except for DS2 in iteration 2, this never corresponded to the optimal value of the parameter.}
\deleted{Typically, the models with the lowest learning rate were those that needed the most rounds to achieve the best performance.}
In the case of DS2 \replaced{in}{of the second} iteration \added{2}, the stopping criterion was not activated, so we extended the maximum rounds to \replaced{2000}{1500} and found an optimal hyperparameter\deleted{s} for \texttt{nrounds} equal to \replaced{1516}{1410}.

The performance of the models obtained by applying XGB to the aforementioned data sets\deleted{to train models} for predicting stellar rotation periods is summarised in \cref{tab:results}.
The number of predictors used to train the models is given under each data set in the header of the table.
The quantities used to assess the quality of the models were described in \cref{sec:performance}.

\begin{table*}[ht]
  \caption{
    Quality assessment of the XGBoost models generated during iterations 1 and 2.
    All quality measures are presented with three significant figures.
    The header indicates the number of predictors in each data set.
    The first two rows show the mean absolute value of the relative residuals measured on the period and frequency regimes;
    the next row corresponds to the interval-based accuracy defined in \cref{sec:performance}, with a width of \qty{10}{\percent} of the reference values;
    the last three rows show the root-mean-squared error and the mean-absolute error in days, both computed on the training and testing sets, and the adjusted coefficient of determination (calculated on the testing set).
    The best results for each metric (except for RMSE and MAE) are shown in bold.
  }
  \label{tab:results}
  \centering
  \resizebox{\linewidth}{!}{
    \begin{tabular}{lccccccccccccccccc}
      \toprule
       & \multicolumn{8}{c}{Iteration 1} & & \multicolumn{8}{c}{Iteration 2} \\
      \cmidrule{2-9}\cmidrule{11-18}
      \begin{tabular}{@{}c@{}}Assessment\\Metric\end{tabular} & \begin{tabular}{@{}c@{}}DS0\\170\end{tabular} & \begin{tabular}{@{}c@{}}DS1\\\repdel{155}{163}\end{tabular} & \begin{tabular}{@{}c@{}}DS2\\\repdel{146}{152}\end{tabular} & \begin{tabular}{@{}c@{}}DS3\\\repdel{146}{152}\end{tabular} & \begin{tabular}{@{}c@{}}DS4\\\repdel{146}{152}\end{tabular} & \begin{tabular}{@{}c@{}}DS5\\\repdel{28}{34}\end{tabular} & \begin{tabular}{@{}c@{}}DS6\\\repdel{28}{34}\end{tabular} & \begin{tabular}{@{}c@{}}DS7\\\repdel{28}{34}\end{tabular} && \begin{tabular}{@{}c@{}}DS0\\170\end{tabular} & \begin{tabular}{@{}c@{}}DS1\\\repdel{155}{163}\end{tabular} & \begin{tabular}{@{}c@{}}DS2\\\repdel{146}{152}\end{tabular} & \begin{tabular}{@{}c@{}}DS3\\\repdel{146}{152}\end{tabular} & \begin{tabular}{@{}c@{}}DS4\\\repdel{146}{152}\end{tabular} & \begin{tabular}{@{}c@{}}DS5\\\repdel{28}{34}\end{tabular} & \begin{tabular}{@{}c@{}}DS6\\\repdel{28}{34}\end{tabular} & \begin{tabular}{@{}c@{}}DS7\\\repdel{28}{34}\end{tabular} \\
      \midrule
      %  & & & & & & & && & & & & & & \\
      \textbf{$\mu_\text{err}$} ($p_\text{rot}$) & \num{0.0535} & \repdel{0.0796}{0.0675} & \repdel{0.0898}{0.0501} & \repdel{0.0617}{0.0486} & \textbf{\repdel{0.0425}{0.0252}} & \repdel{0.0549}{0.0260} & \repdel{0.0452}{0.0229} & \repdel{0.0760}{0.0328} && 0.0423 & \repdel{0.0775}{0.0628} & \repdel{0.0849}{0.0517} & \repdel{0.0607}{0.0435} & \textbf{\repdel{0.0424}{0.0259}} & \repdel{0.0560}{0.0247} & \repdel{0.0458}{0.0221} & \repdel{0.0722}{0.0316} \\
      \textbf{$\mu_\text{err}$} ($f_\text{rot}$) & \num{0.0364} & \repdel{0.0549}{0.0379} & \repdel{0.0556}{0.0335} & \repdel{0.0475}{0.0302} & \textbf{\repdel{0.0409}{0.0243}} & \repdel{0.0480}{0.0496} & \repdel{0.0427}{0.0223} & \repdel{0.0532}{0.0302} && 0.0300 & \repdel{0.0549}{0.0302} & \repdel{0.0528}{0.0331} & \repdel{0.0463}{0.0277} & \textbf{\repdel{0.0405}{0.0242}} & \repdel{0.0489}{0.0233} & \repdel{0.0425}{0.0213} & \repdel{0.0499}{0.0289} \\
      \addlinespace[2mm]
      \textbf{acc$_\text{10}$} & 0.923 & \repdel{0.859}{0.926} & \repdel{0.854}{0.923} & \repdel{0.889}{0.938} & \textbf{\repdel{0.896}{0.945}} & \repdel{0.884}{0.948} & \repdel{0.892}{0.953} & \repdel{0.861}{0.928} && 0.931 & \repdel{0.861}{0.935} & \repdel{0.868}{0.927} & \textbf{\repdel{0.897}{0.947}} & \repdel{0.896}{0.944} & \repdel{0.884}{0.951} & \repdel{0.895}{0.954} & \repdel{0.873}{0.929} \\
      \addlinespace[2mm]
      \textbf{RMSE} (train/test) & \begin{tabular}{@{}c@{}}0.872\\2.22\end{tabular} & \begin{tabular}{@{}c@{}}\repdel{0.741}{0.806}\\\repdel{2.66}{2.19}\end{tabular} & \begin{tabular}{@{}c@{}}\repdel{0.686}{0.714}\\\repdel{2.84}{2.61}\end{tabular} & \begin{tabular}{@{}c@{}}\repdel{0.551}{0.713}\\\repdel{2.13}{1.83}\end{tabular} & \begin{tabular}{@{}c@{}}\repdel{0.619}{0.447}\\\repdel{2.08}{1.82}\end{tabular} & \begin{tabular}{@{}c@{}}\repdel{0.629}{0.780}\\\repdel{2.18}{1.41}\end{tabular} & \begin{tabular}{@{}c@{}}\repdel{0.584}{0.589}\\\repdel{2.307}{1.52}\end{tabular} & \begin{tabular}{@{}c@{}}\repdel{0.979}{0.630}\\\repdel{2.69}{2.11}\end{tabular} && \begin{tabular}{@{}c@{}}0.547\\2.10\end{tabular} & \begin{tabular}{@{}c@{}}\repdel{0.752}{0.678}\\\repdel{2.66}{2.18}\end{tabular} & \begin{tabular}{@{}c@{}}\repdel{0.574}{0.638}\\\repdel{2.84}{2.56}\end{tabular} & \begin{tabular}{@{}c@{}}\repdel{0.467}{0.484}\\\repdel{2.09}{1.79}\end{tabular} & \begin{tabular}{@{}c@{}}\repdel{0.511}{0.478}\\\repdel{2.10}{1.76}\end{tabular} & \begin{tabular}{@{}c@{}}\repdel{0.590}{0.551}\\\repdel{2.17}{1.40}\end{tabular} & \begin{tabular}{@{}c@{}}\repdel{0.393}{0.378}\\\repdel{2.33}{1.49}\end{tabular} & \begin{tabular}{@{}c@{}}\repdel{0.650}{0.767}\\\repdel{2.65}{2.02}\end{tabular} \\
      \addlinespace[2mm]
      \textbf{MAE} (train/test) & \begin{tabular}{@{}c@{}}0.377\\0.803\end{tabular} & \begin{tabular}{@{}c@{}}\repdel{0.452}{0.370}\\\repdel{1.27}{0.789}\end{tabular} & \begin{tabular}{@{}c@{}}\repdel{0.425}{0.332}\\\repdel{1.34}{0.877}\end{tabular} & \begin{tabular}{@{}c@{}}\repdel{0.339}{0.328}\\\repdel{1.03}{0.601}\end{tabular} & \begin{tabular}{@{}c@{}}\repdel{0.384}{0.233}\\\repdel{1.07}{0.630}\end{tabular} & \begin{tabular}{@{}c@{}}\repdel{0.378}{0.340}\\\repdel{1.08}{0.531}\end{tabular} & \begin{tabular}{@{}c@{}}\repdel{0.357}{0.267}\\\repdel{1.13}{0.577}\end{tabular} & \begin{tabular}{@{}c@{}}\repdel{0.575}{0.301}\\\repdel{1.28}{0.775}\end{tabular} && \begin{tabular}{@{}c@{}}0.241\\0.719\end{tabular} & \begin{tabular}{@{}c@{}}\repdel{0.453}{0.273}\\\repdel{1.29}{0.830}\end{tabular} & \begin{tabular}{@{}c@{}}\repdel{0.351}{0.279}\\\repdel{1.28}{0.821}\end{tabular} & \begin{tabular}{@{}c@{}}\repdel{0.289}{0.229}\\\repdel{0.995}{0.550}\end{tabular} & \begin{tabular}{@{}c@{}}\repdel{0.321}{0.237}\\\repdel{1.07}{0.628}\end{tabular} & \begin{tabular}{@{}c@{}}\repdel{0.345}{0.252}\\\repdel{1.08}{0.510}\end{tabular} & \begin{tabular}{@{}c@{}}\repdel{0.251}{0.181}\\\repdel{1.12}{0.553}\end{tabular} & \begin{tabular}{@{}c@{}}\repdel{0.395}{0.326}\\\repdel{1.22}{0.728}\end{tabular} \\
      \addlinespace[2mm]
      \textbf{R$^2_\text{adj}$} (test) & 0.963 & \repdel{0.947}{0.965} & \repdel{0.942}{0.949} & \textbf{\repdel{0.955}{0.968}} & \repdel{0.948}{0.959} & \repdel{0.952}{0.980} & \repdel{0.938}{0.972} & \repdel{0.943}{0.968} && 0.967 & \repdel{0.947}{0.965} & \repdel{0.942}{0.950} & \textbf{\repdel{0.957}{0.969}} & \repdel{0.948}{0.962} & \repdel{0.953}{0.980} & \repdel{0.937}{0.973} & \repdel{0.945}{0.973} \\
      \bottomrule
    \end{tabular}
  }
\end{table*}

Overall, the XGBoost models were wrong, on average, approximately between \replaced{4.1}{2.2} and \replaced{\qty{9.0}{\percent}}{\qty{6.8}{\percent}} of the time in iteration 1, and between \replaced{4.1}{2.1} and \replaced{\qty{8.5}{\percent}}{\qty{6.3}{\percent}} of the time in iteration 2, as indicated by $\mu_\text{err}$.
DS\replaced{4}{6} was the least likely to make an error, and DS5 had the best adjusted R-squared.
\deleted{The highest average relative error of DS5 in iteration 1, measured in frequency domain, is due to the scaling at rotation periods close to zero.}

\begin{figure*}[!ht]
  \begin{center}
    \includegraphics[width=0.99\linewidth]{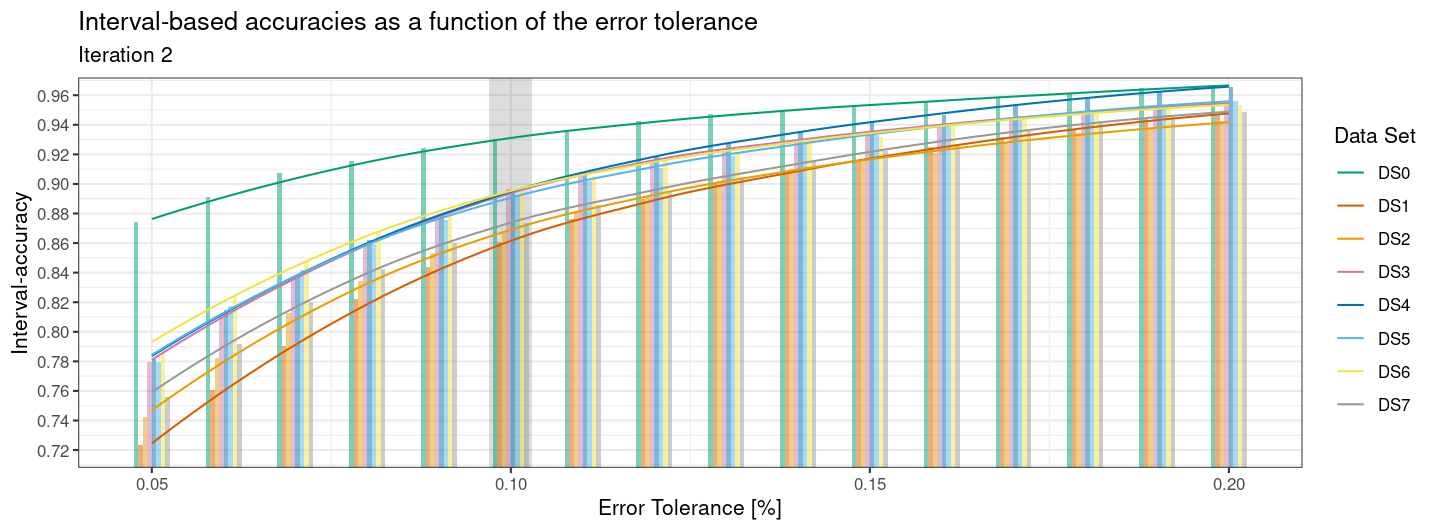}
    \caption{
      Evolution of the interval-based accuracies with the error tolerance.
      The models are indicated in different colours.
      The lines correspond to a Locally Weighted Scatterplot Smoothing (LOESS) regression of the accuracies for each model.
      The \qty{10}{\percent}-accuracy is highlighted by the vertical gray-shaded column in the background.
      The model with the highest overall accuracy was DS6.
    }
    \label{fig:acc-iter2}
  \end{center}
\end{figure*}

\replaced{Excluding DS0, the}{The} \qty{10}{\percent}-accuracy varied between \replaced{85.4}{92.3} and \replaced{\qty{89.6}{\percent}}{\qty{95.3}{\percent}} in iteration 1, and between \replaced{86.1}{92.7} and \replaced{\qty{89.7}{\percent}}{\qty{95.4}{\percent}} in iteration 2, with an increase of about 2 to \replaced{4}{3} points when stars with rotation periods \added{shorter than \qty{7}{\day} and} greater than \qty{45}{\day} were removed and the number of predictors was reduced to the \replaced{28}{34} most important.
\replaced{In the case of}{The exception was} DS7, \replaced{there is also an increase compared to DS1 and DS2. However, it is less significant than in the other cases.}{which achieved a similar accuracy to DS2 in the second iteration.}
Therefore, the \qty{10}{\percent}-accuracy seems to increase with smaller data sets and especially with an improvement in data quality.
The evolution of the interval-based accuracy with the error tolerance for the models obtained during iteration 2 is illustrated in \cref{fig:acc-iter2}.
Overall, \added{DS0 stands out from the other models, and} the interval-based accuracy increases with the error tolerance, \ie when we are willing to accept larger errors.
We find that
\begin{enumerate*}[label=(\roman*), font=\itshape]
  \item \added{upon the exclusion of DS0,} it is possible to identify two \added{distinct} groups of models\deleted{, mainly in the lowest error tolerance regime}: the first, composed of\deleted{DS0,} DS1, DS2, and DS7, with lower accuracies; and the second, containing the remaining models, with higher interval-based accuracies;
  \item \added{particularly within the lowest error tolerance regime,} there is \replaced{a discernible enhancement}{a slight improvement} in the accuracies of \replaced{DS2}{DS1} over \replaced{DS1}{DS0} \added{and DS7 over DS2}, \ie when we \replaced{run an initial selection of variables, as described in \cref{sec:data-engineering}, and reduce the set of predictors to the most relevant ones}{remove the rotation period predictors, especially for error tolerances above \qty{7}{\percent}};
  \item overall, \replaced{DS1}{DS2} is the model with the worst accuracy;%
  \deleted{\item DS1 always performs better than DS2, \ie there is no apparent performance advantage to running an initial selection of variables as described in \cref{sec:data-engineering} beyond the natural reduction in training time;}%
  %%% replace the previous line by the following one if compiling with the
  %%% switch \showchangestrue on
  % \item\deleted{DS1 always performs better than DS2, \ie there is no apparent performance advantage to running an initial selection of variables as described in \cref{sec:data-engineering} beyond the natural reduction in training time;}%
  \item \added{in the highest tolerance regime, DS4 reaches values that are comparable to those of DS0;}
  \item DS7 \replaced{consistently outperforms DS1 and DS2}{performs similarly to DS1, except for error tolerances between 7 and \qty{9}{\percent}, where its interval-based accuracy is about 1 point lower};
  \item there is a noticeable increase in performance when potentially problematic targets, \ie stars with rotation periods above \qty{45}{\day} and below \qty{7}{\day}, are removed, especially for lower error tolerances; and
  \item the performance increases further when only the \replaced{28}{34} most important predictors are used to train the models, with the interval-based accuracy always remaining above \replaced{\qty{87}{\percent}}{\qty{90}{\percent}}.
\end{enumerate*}

The adjusted-R$^2$ measured on the testing set was always equal to or greater than \replaced{\qty{94}{\percent}}{\qty{95}{\percent}}.
The model with the best goodness of fit measured by this metric was \replaced{DS3}{DS5} with \replaced{\qty{96}{\percent}}{\qty{98}{\percent}}, followed by \replaced{DS5}{DS6 and DS7} with \replaced{\qty{95}{\percent}}{\qty{97}{\percent}}.
These values seem to indicate a slightly greater ability of the predictors to explain the variability of the response when stars with short rotation periods (less than \qty{7}{\day}) are \added{kept in the data set} used to train the model.
When compared with the corresponding unadjusted coefficients of determination, these $R_\text{adj}^2$ values were all of the same magnitude within each model, with differences typically to the third decimal place.

\begin{figure}[!ht]
  \begin{center}
    \includegraphics[width=\linewidth]{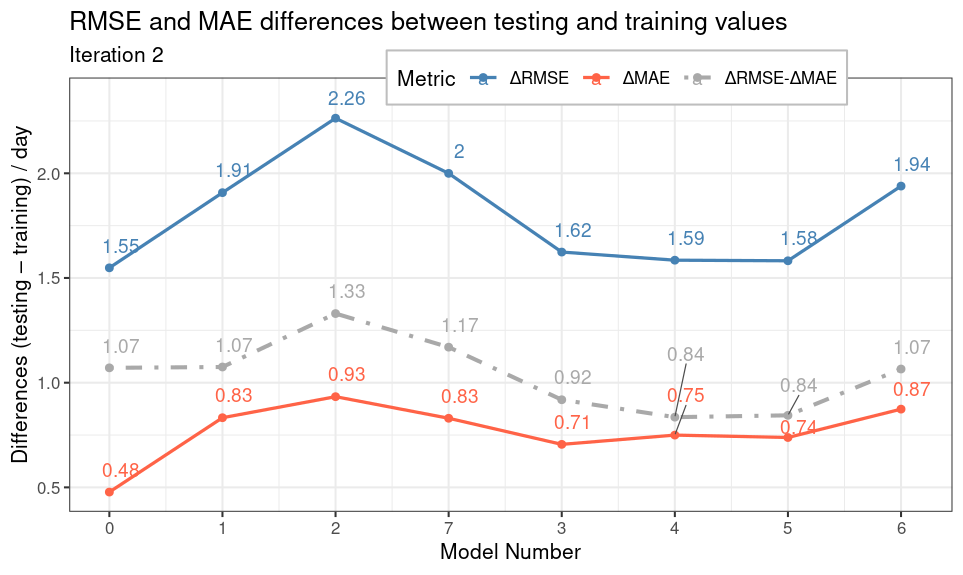}
    \caption{
      RMSE and MAE differences---respectively $\Delta$RMSE (blue thick line) and $\Delta$MAE (red thin line)---between the testing and training sets for the models generated in iteration\deleted{s 1 and} 2.
      For both the RMSE and MAE, the differences are highest at DS2 and lowest at \replaced{DS0}{DS5}.
      The dotdashed grey line indicates $\Delta\text{RMSE} - \Delta\text{MAE}$, which is\deleted{also} maximum at DS2 and minimum at \added{DS4 and} DS5.
    }
    \label{fig:rmse-mae}
  \end{center}
\end{figure}

\Cref{fig:rmse-mae} shows the RMSE and MAE differences between the testing and training sets---$\Delta\text{RMSE}$ (blue thick line) and $\Delta\text{MAE}$ (red thin line), respectively---and the difference between the two, $\Delta\text{RMSE} - \Delta\text{MAE}$ (dotdashed grey line), for all the models generated in\deleted{both} iteration\deleted{s} 2.
\added{It can be observed that} \replaced{b}{B}oth\deleted{the} differences\replaced{, $\Delta\text{RMSE}$ and $\Delta\text{MAE}$,}{between the testing and training RMSE and between the testing and training MAE} \added{reach a minimum value with DS0;}
\added{following an initial increase in DS1 and DS2, a} \replaced{tendency towards a}{tend to} decrease \added{can be identified} as the \added{data} sets get smaller and the quality of the data increases.
However, even if the downward trend continues, these differences reach values greater than the previous models in the case of DS2\deleted{DS4,} and DS6.\added{\footnote{\added{There is a marginal increase in $\Delta\text{MAE}$ in DS4 compared to DS3.}}}
The general downward trend in the solid lines indicate\added{s} that\added{, with the exception of DS6,} the degree of overfitting \replaced{is less pronounced}{decreases} with the number of the model, \ie with\deleted{smaller} data sets \added{comprising a smaller number but more relevant predictors} and higher data quality.
In iteration \replaced{2}{1}, $\Delta\text{RMSE}$ varied between \replaced{\qty{1.55}{\day}}{\qty{1.12}{\day}} \added{for DS0} and \replaced{\qty{2.26}{\day}}{\qty{1.89}{\day}} \added{for DS2.} \deleted{for the first five models together with DS7, and then dropped below \qty{1}{\day} for DS5 and DS6.}%
\added{The latter was, therefore, the model with the highest degree of overfitting.}
\added{After DS0,} \deleted{It reached the minimum value of \qty{0.63}{\day} for} DS5\deleted{, indicating that this} was the model with the least overfitting.
\deleted{The highest degree of overfitting occurred for DS2.}
\replaced{A similar}{This} behaviour was \deleted{also} observed in the case of $\Delta\text{MAE}$\added{, where it reached the minimum value of \qty{0.48}{\day} for DS0 and a maximum of \qty{0.93}{\day} for DS2}.
The difference $\Delta\text{RMSE} - \Delta\text{MAE}$ suggests that, with the exception of DS2 \added{and DS6}, the variance of individual errors generally decreased with smaller data sets\deleted{, and indicates that with the exception of DS4 and DS7, the overfitting increased slightly in iteration 2 compared to iteration 1}.
This greater overfitting did not prevent the models from performing slightly better in iteration 2 \added{than in iteration 1}.
We can also see that \added{the overfitting is less pronounced in} DS7\deleted{overfits less} than \replaced{in DS2}{DS0} and \added{is similar to} DS1\deleted{ in the second iteration}.
\replaced{The evolution of these differences is in line with the behaviour of}{The increase in these differences in RMSE and MAE in the DS2, DS4 and DS6 models parallels the decrease in} the adjusted R$^2$, \added{and} \deleted{relative to their predecessor models.}
\deleted{This} seems to indicate that the removal of rotation periods shorter than seven days \added{and larger than 45 days} has a \replaced{positive}{negative} impact on the performance of the models.

\added{
  The mean values of the standard deviations for each model, $\bar\sigma$, and the ratio between these uncertainties and the mean values of the predictions carried out on the testing set, $\bar\sigma / \bar{\hat{p}}_\text{rot}$, estimated in uniform bins of unit length centred on integer values, are indicated in \cref{tab:error-bars}.
  For each prediction, the associated error bar is taken to be equal to the standard deviation in the corresponding bin.
  The values are indicative of the variability of the predictions.
  The mean length of the error bars oscillates between 0.28 and \qty{0.29}{\day}, while the mean relative uncertainty is approximately equal to \qty{1.2}{\percent}, ranging between 0.0115 and 0.0122.
}

\begin{table*}[ht]
  \caption{
    \added{Mean values of the error bars, in days, and ratio between the uncertainty and the average predicted values, for each model.}
  }
  \label{tab:error-bars}
  \centering
  % \resizebox{\linewidth}{!}{
    \begin{tabular}{lcccccccc}
      \toprule
       & DS0 & DS1 & DS2 & DS3 & DS4 & DS5 & DS6 & DS7 \\
      \midrule
      $\bar\sigma$ / day & 0.284 & 0.288 & 0.290 & 0.286 & 0.285 & 0.287 & 0.283 & 0.286 \\
      $\bar\sigma / \bar{\hat{p}}_\text{rot}$ & 0.0116 & 0.0118 & 0.0119 & 0.0123 & 0.0115 & 0.0122 & 0.0115 & 0.0117 \\
      \bottomrule
    \end{tabular}
  % }
\end{table*}

\begin{figure}[!ht]
  \centering
  \includegraphics[width=0.99\linewidth]{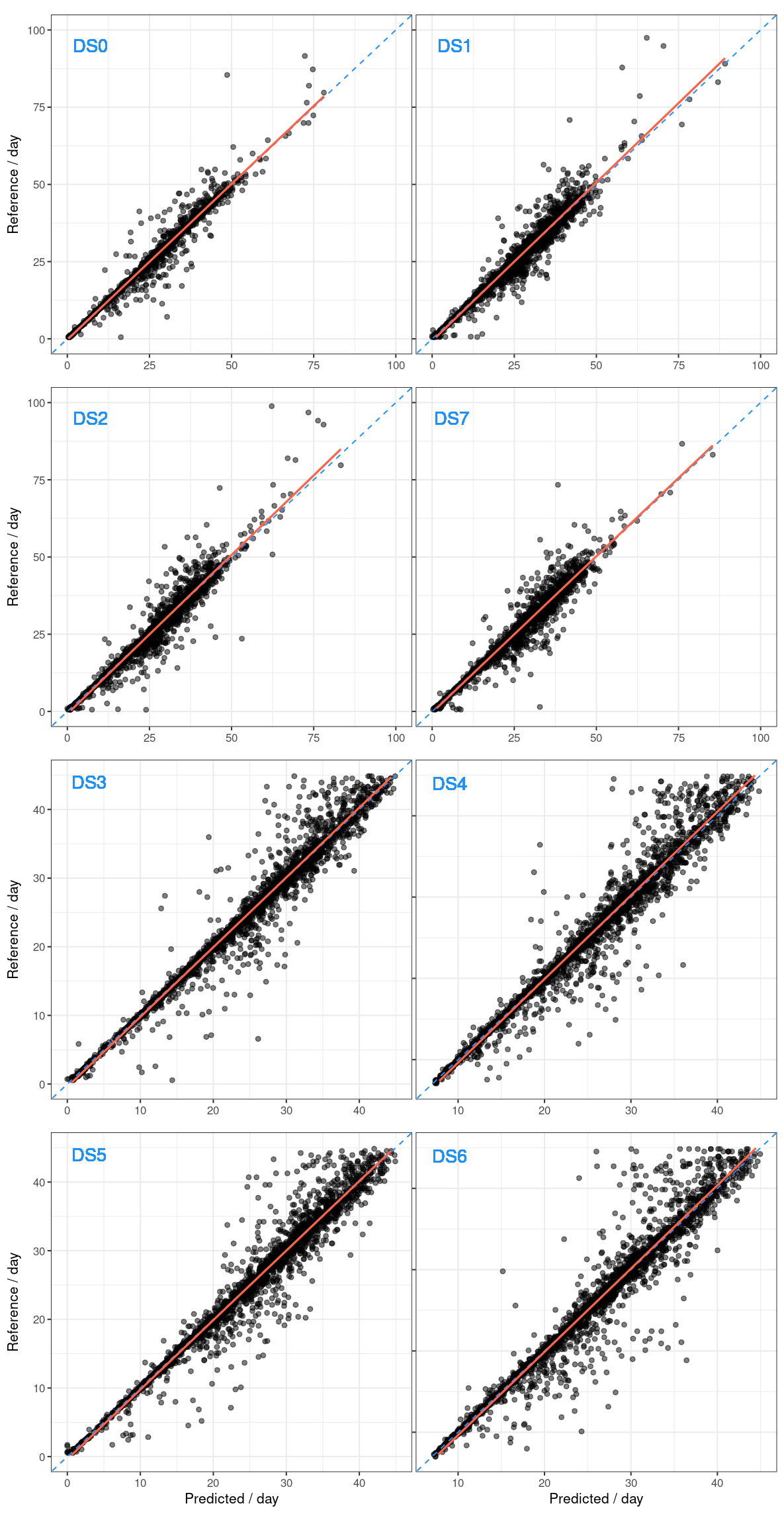}
  \caption{
    Scatter plots of the reference rotation period as a function of the predicted values for the models built in iteration 2.
    The blue dashed lines indicate the identity function, and the red solid lines represent the linear model between the ground truth and the predictions.
    DS7 has been placed next to DS1 and DS2 to facilitate comparison of the models.
  }
  \label{fig:scatter-grid}
\end{figure}

The scatter plots of the ground truth against the predicted values for all the models built during iteration 2 are shown in \cref{fig:scatter-grid} and, with additional marginal density plots, in \cref{fig:scatter-density-grid}.
The blue dashed lines indicate the identity function, and the red solid lines represent the linear models between the predicted and actual values.
\added{In order to enhance the clarity of the plot, the error bars associated with the predictions have been omitted.}
Most of the points fall on \added{or near} the $y = x$ line, but some outliers can be seen, representing both under- and over-predicted cases.
Nevertheless, all graphs show a high degree of positive correlation between the predicted and the reference values of the stellar rotation periods, indicating that there is generally good agreement between the predictions and the ground truth.
This is confirmed by the marginal histograms and density plots for both the predicted and reference values in each panel of \cref{fig:scatter-density-grid}, which are similar in terms of centrality, dispersion, and kurtosis.
The dispersion of the observations relative to the identity line is not significantly large in all the models, but it increases with the predictions, \ie it has different values for low and high values of the response---an indication of heteroscedasticity.
The R-squared reported in the panels correspond to the coefficient of determination between the predicted and the reference values.
As expected, they are not significantly different from the R$^2_\text{adj}$ reported in \cref{tab:results}, given the large total number of observations, $n$.
The F-statistics shown in the graphs result from testing the null hypothesis that all of the regression coefficients are equal to zero.
Overall, the F-tests and the corresponding $p$-values considered in the panels indicate that the sets of independent variables are jointly significant.
Models DS0, DS1, DS2, and DS7 have similar performance for rotation periods shorter than about \qty{50}{\day}.
If we consider rotation periods longer than \qty{50}{\day}, the dispersion is lowest for DS7.
\deleted{DS0 was not able to make predictions for rotation periods longer than about \qty{77}{\day}.}

\begin{figure}[!htp]
  \centering
  \includegraphics[width=0.99\linewidth]{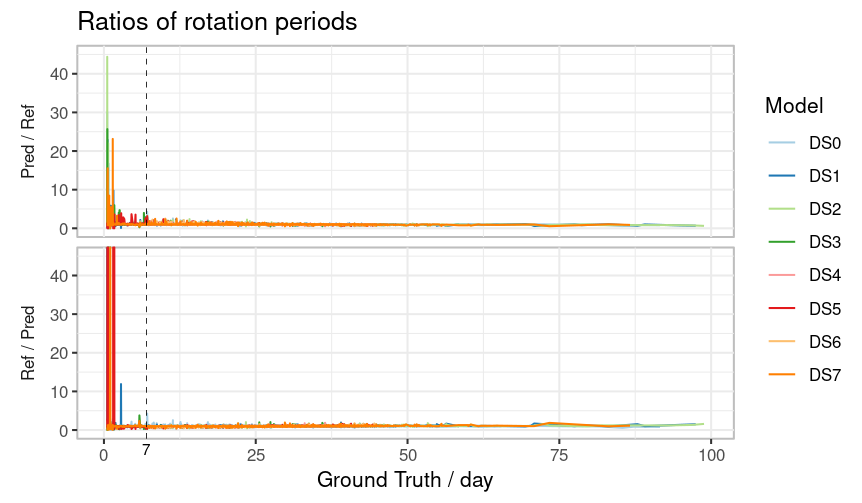}
  \caption{
    Predictions over the ground truth (top panel) and its inverse (bottom panel) for the models obtained in iteration 2.
    The top plot highlights the magnitude of over-predictions, while the bottom plot highlights the amplitude of under-predictions.
  }
  \label{fig:ratios}
\end{figure}

In \cref{fig:ratios}, we have plotted the ratio between the predicted and the reference values (top panel), and the ratio between the ground truth and the predictions (bottom panel), both against the reference values.
The former highlights the over-predictions, while the latter emphasises the under-predictions.
The highest over- and under-prediction ratios are equal to \replaced{44.4}{40.9} and \replaced{30.2}{2.86}, respectively.
There are \replaced{nine}{three} under-prediction ratios equal to infinity, corresponding to zero values predicted by the model (we made negative predictions equal to zero).
The largest ratios occur only for predictions close to zero, \ie for very small rotation periods, typically below \qty{7}{\day}, corresponding to fast rotators.
\added{It can be seen that} \replaced{a}{A}bove the \qty{7}{\day} threshold, \added{both} the overestimations \added{and the underestimations} are \replaced{at most about four times}{always less than twice} the reference values.
\deleted{The range of the $y$-axis is smaller in the lower panel.}
\deleted{It can be seen that above the \qty{7}{\day} threshold, the underestimations are at most about half of the reference value.}
\deleted{The better performance of DS7 in the slower rotation regime compared to models DS0, DS1 and DS2 is evident from the plot in the lower panel.}

\begin{figure}[!htbp]
  \begin{center}
    \includegraphics[width=\linewidth]{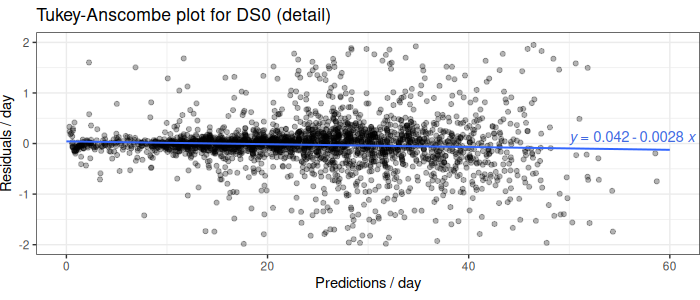}
    \caption{
      Detail of the Tukey-Anscombe plot for DS0 during iteration 2, where we have restricted the $y$-axis to the values $-2$ and $2$.
      The plot shows no meaningful slope, but it is possible to identify some degree of heteroscedasticity.}
    \label{fig:ta-ds0}
  \end{center}
\end{figure}

The Tukey-Anscombe (TA) plots of the raw residuals and of the \qty{10}{\percent}-error are shown in \cref{fig:residuals}.
For each model, the $y$-axis in the top panel corresponds to the residuals calculated on the rotation periods (in days), and in the bottom panel to the interval-based error calculated on the rotation frequencies (in $\text{days}^{-1}$).
The $x$-coordinate always represents the predictions in days.
In all the plots, the points fluctuate randomly around the horizontal line passing through zero, forming a cloud that is approximately symmetrical around it.
However, this cloud is not rectangular in shape.%
\begin{figure}[ht]
  \centering
  \includegraphics[width=\linewidth]{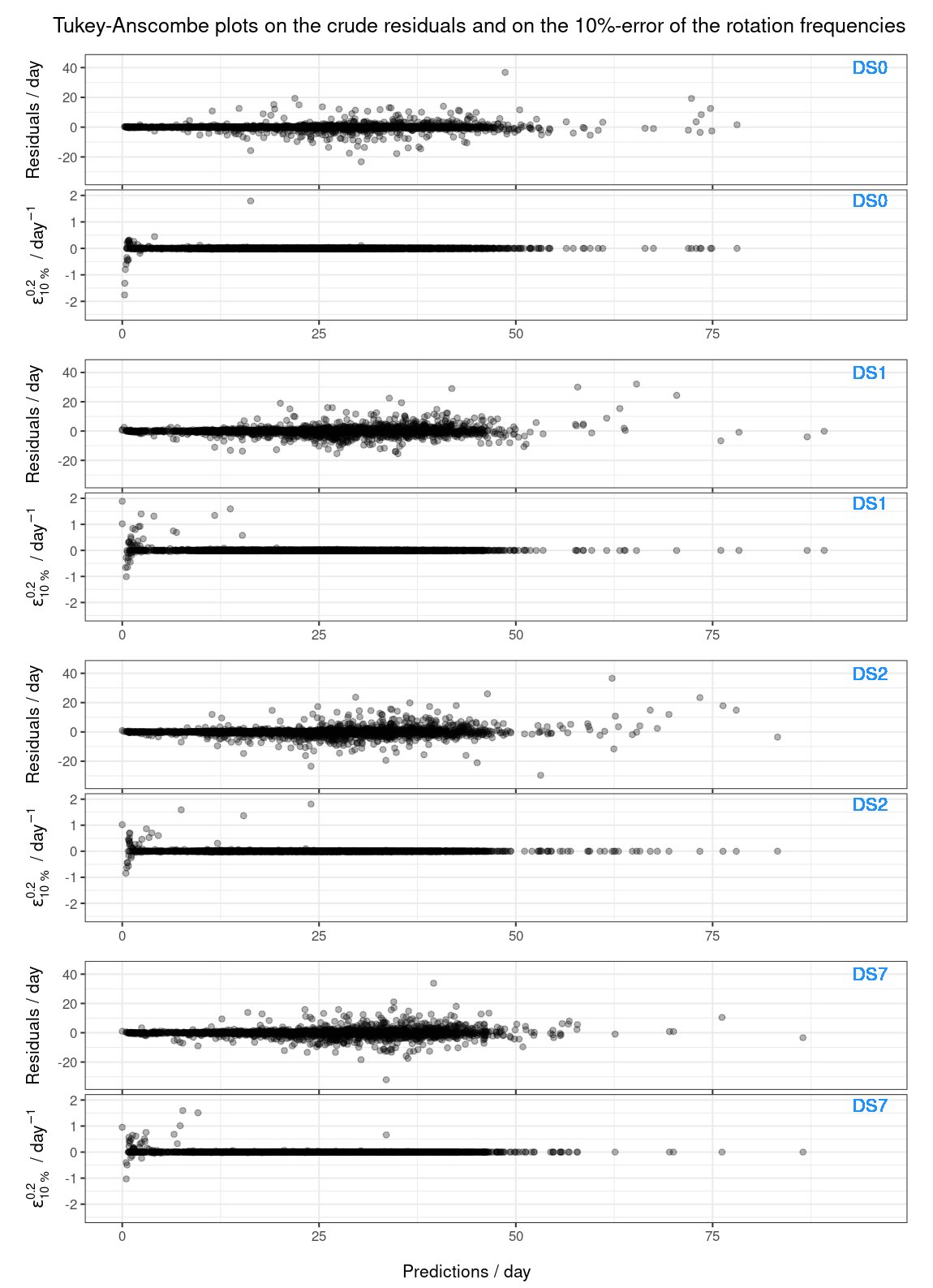}
  \caption{
    Tukey-Anscombe plots
    \begin{enumerate*}[label=(\roman*), font=\itshape]
      \item of the crude residuals computed on the rotation periods (top panels) and
      \item of the \qty{10}{\percent}-error computed on the rotation frequencies (bottom panels)
    \end{enumerate*}
    for the models constructed during iteration 2 (continues on the next column).
  }
  \label{fig:residuals}
\end{figure}
\begin{figure}[ht]
  \ContinuedFloat
  \centering
  \includegraphics[width=\linewidth]{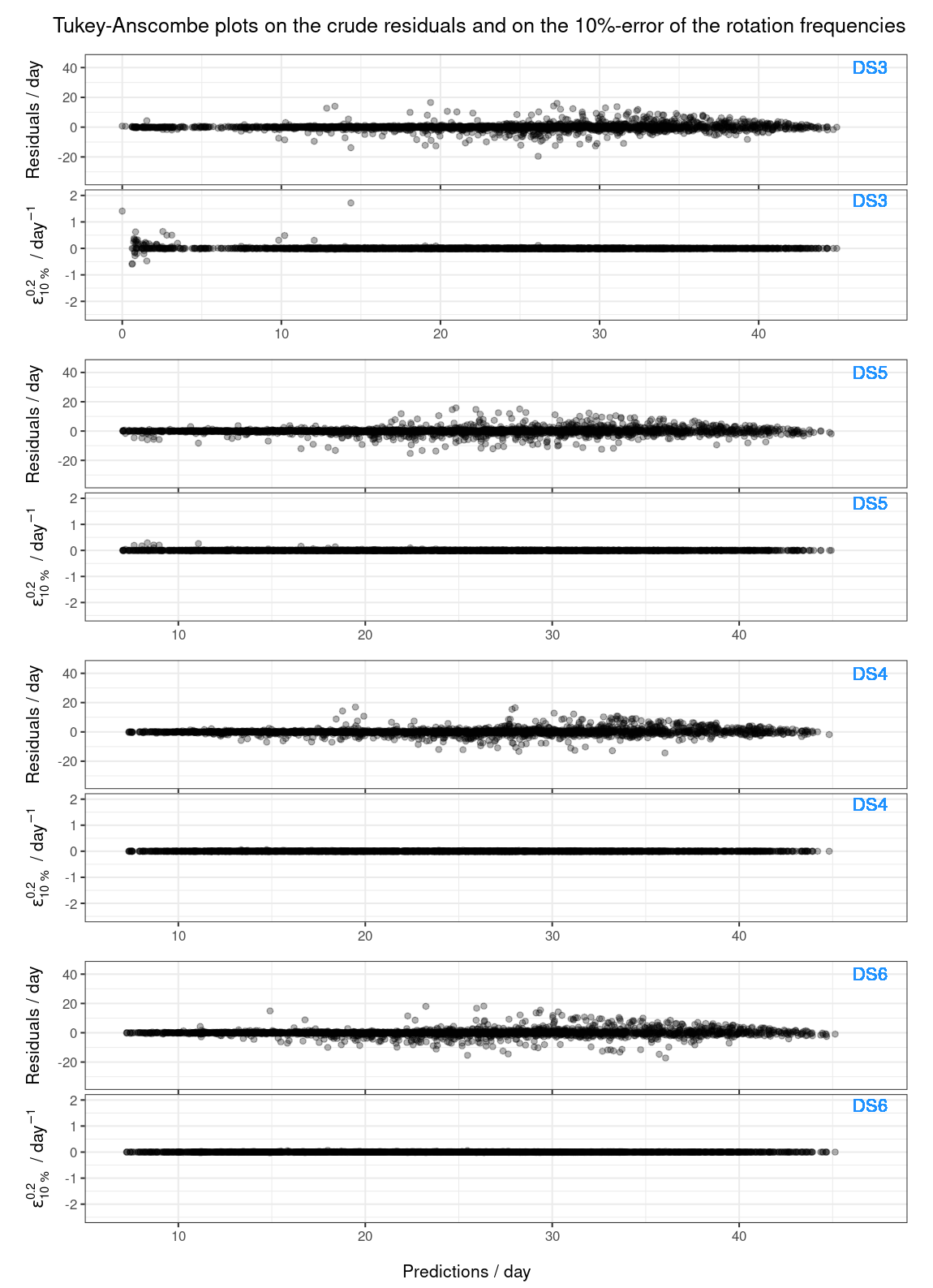}
  \caption{
    Tukey-Anscombe plots
    \begin{enumerate*}[label=(\roman*), font=\itshape]
      \item of the crude residuals computed on the rotation periods (top panels) and
      \item of the \qty{10}{\percent}-error computed on the rotation frequencies (bottom panels)
    \end{enumerate*}
    for the models constructed during iteration 2 (continued).
  }
  \label{fig:residuals-2}
\end{figure}
A detail of the TA plot for DS0, where we have constrained the $y$-axis to vary between $-2$ and 2, is shown in \cref{fig:ta-ds0}.
We have added a regression line to the graph, which does not show a meaningful slope.
The variance of the residuals increases with the predictions, indicating some degree of heteroscedasticity.
All models share this characteristic of the TA plot.

The most important predictors for the models obtained in iteration 2 are highlighted in the bar plots of \cref{fig:importance}.
Apart from DS0, the variables belonging to the CS and GWPS families typically contributed the most to the models.
In the case of DS0 (the only data set where they were present), the rotation periods were naturally the dominant variables.
Having obtained DS1, DS2, DS3 and DS4, we calculated the union of the 20 most important variables for these models.
We were left with a set of \replaced{28}{34} predictors, which we used to create the DS5, DS6, and DS7 data sets and their respective models.
These variables were the following:
%
% \begin{multicols}{2}
  \begin{itemize}
    \item \added{\texttt{cs\_chiq\_55}}
    \item \texttt{cs\_gauss\_1\_1\_20}
    \item \texttt{cs\_gauss\_1\_2\_20}
    \item \texttt{cs\_gauss\_2\_2\_20}
    \item \texttt{cs\_gauss\_2\_2\_55}
    \item \texttt{cs\_gauss\_3\_1\_20}
    \item \texttt{cs\_gauss\_3\_1\_55}
    \item \added{\texttt{cs\_gauss\_3\_1\_80}}
    \item \added{\texttt{cs\_gauss\_3\_2\_55}}
    \item \added{\texttt{f\_07}}
    \item \texttt{gwps\_gauss\_1\_1\_20}
    \item \texttt{gwps\_gauss\_1\_2\_20}
    \item \texttt{gwps\_gauss\_1\_4\_55}
    \item \texttt{gwps\_gauss\_2\_2\_20}
    \item \texttt{gwps\_gauss\_2\_2\_55}
    \item \texttt{gwps\_gauss\_2\_2\_80}
    \item \added{\texttt{gwps\_gauss\_3\_1\_20}}
    \item \added{\texttt{gwps\_gauss\_3\_1\_55}}
    \item \added{\texttt{gwps\_gauss\_3\_1\_80}}
    \item \texttt{gwps\_gauss\_3\_2\_20}
    \item \texttt{h\_acf\_55}
    \item \texttt{h\_cs\_20}
    \item \added{\texttt{length}}
    \item \texttt{sph}
    \item \texttt{sph\_acf\_err\_20}
    \item \texttt{sph\_cs\_err\_20}
    \item \texttt{sph\_gwps\_20}
    \item \texttt{teff}
  \end{itemize}
% \end{multicols}
%
\replaced{The predictors}{Of these 34, the 30} that contributed most to the models are shown on the $y$-axis of \cref{fig:importance}.
\added{For models DS0 to DS4, we show the 30 most relevant variables for training;
for the rest of the models, the plots show the 28 most important predictors mentioned above.}
We emphasise that Astro variables, such as the mass and the effective temperature, \added{although they had some relevance for DS0, DS1, and DS2, they} contribute little to the \added{DS3 and DS4} models.
TS-type variables, such as the photometric activity proxy, have some relevance and are important to support the model---we observe a drop in model performance when they are removed.
\begin{figure}[!ht]
  \begin{center}
    \includegraphics[width=\linewidth]{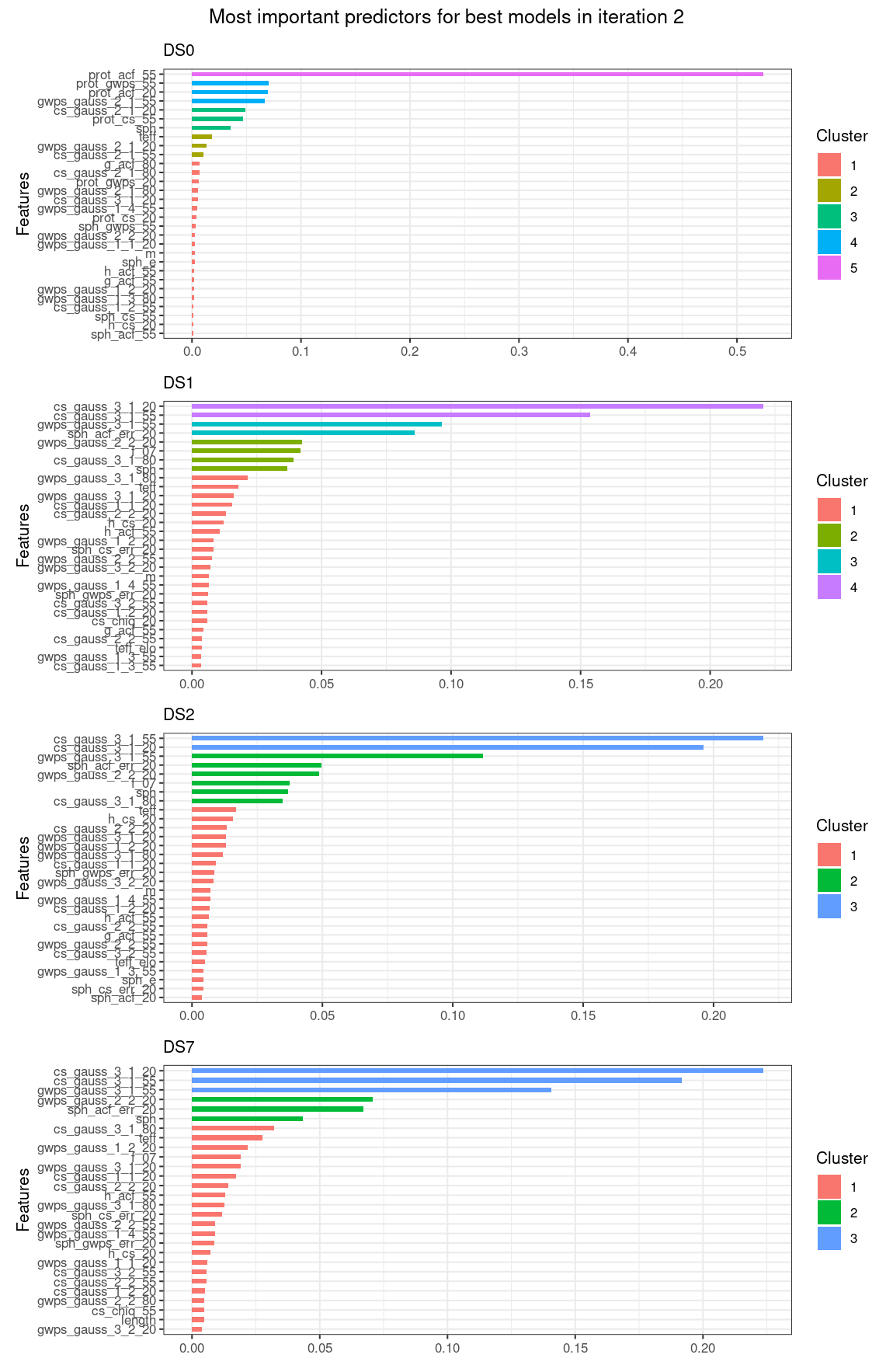}
    \caption{
      Variable importance for the best models generated in iteration 2.
      In all models except DS0, the most important predictor was \texttt{cs\_gauss\_2\_1\_55}, followed by \texttt{gwps\_gauss\_2\_1\_55} (continues on the next column).
    }
    \label{fig:importance}
  \end{center}
\end{figure}
\begin{figure}[!ht]
  \ContinuedFloat
  \begin{center}
    \includegraphics[width=\linewidth]{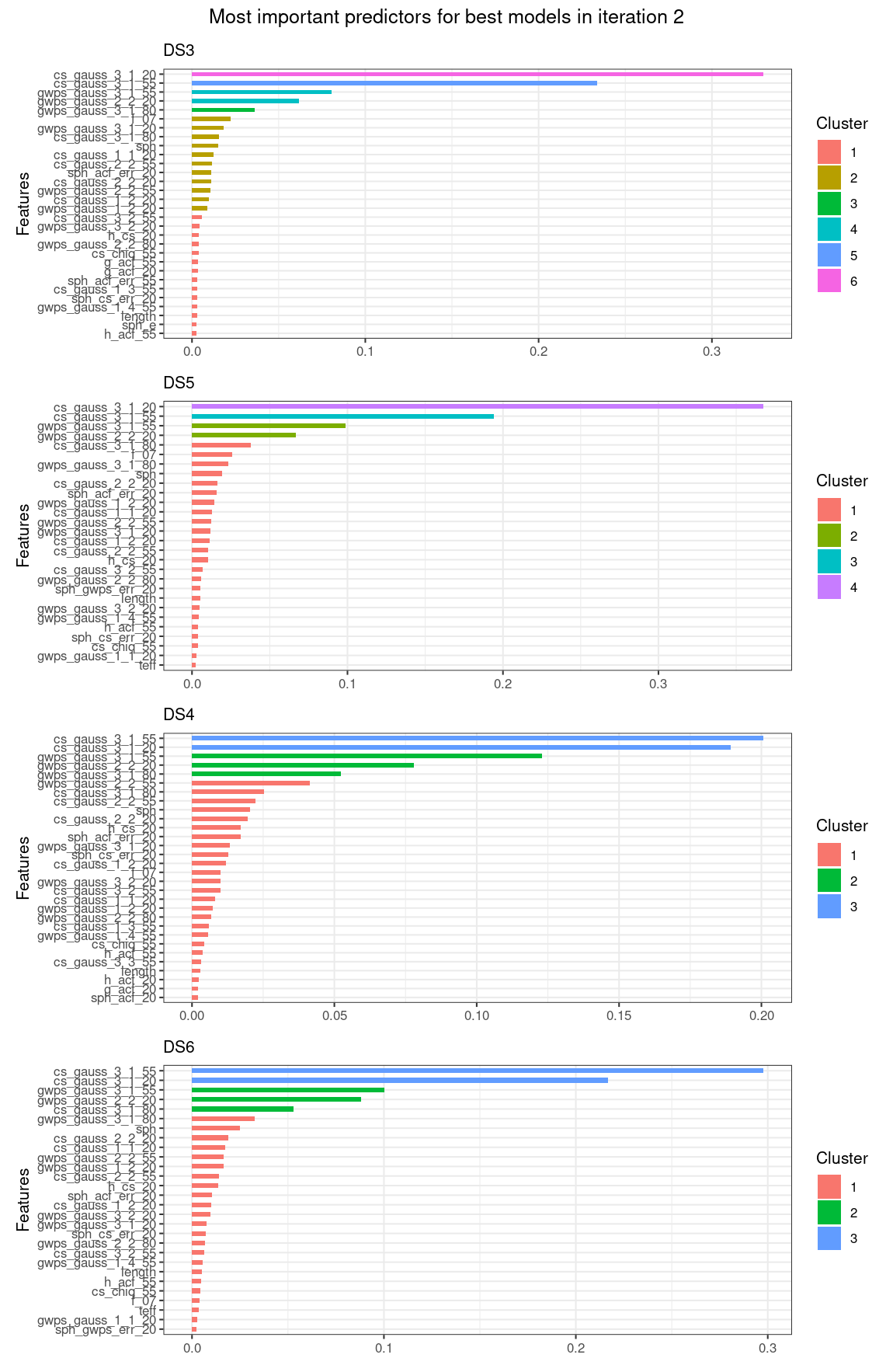}
    \caption{
      Variable importance for the best models generated in iteration 2.
      In all models except DS0, the most important predictor was \texttt{cs\_gauss\_2\_1\_55}, followed by \texttt{gwps\_gauss\_2\_1\_55} (continued).
    }
    \label{fig:importance-2}
  \end{center}
\end{figure}
%
%%%%%%%%%%%%%%%%%%%%%%%%%%%%%%%%%%%%%%%%%%%%%%%%%%%%%%%%%%%%%%%%%%%%%%%%%%%%%%%%

\section{Discussion}
\label{sec:discussion}
The main findings from this work are highlighted in \cref{fig:acc-iter2,fig:scatter-grid,fig:importance}.
On the basis of the latter, \added{if DS0 is excluded,} the results show that the \replaced{three}{two} most important variables for predicting rotation periods of K and M stars from the \textit{Kepler} catalogue arguably belong to the Composite Spectrum and the Global Wavelet Power Spectrum groups, and specifically are \replaced{\texttt{cs\_gauss\_3\_1\_20}, \texttt{cs\_gauss\_3\_1\_55}, and \texttt{gwps\_gauss\_3\_1\_55}}{\texttt{cs\_gauss\_2\_1\_55} and \texttt{gwps\_gauss\_2\_1\_55}}.
After removing\deleted{the Prot} features \added{containing rotation period values} from the data set, these variables were always in the top 10 in terms of importance and stood out from the rest of the features.
They correspond to the \replaced{standard deviation}{central period} of the first Gaussian fitted to the CS and GWPS for the \added{20- and} 55-day filter\added{s}, respectively.
Not only these \replaced{three}{two} predictors, but most of the CS and GWPS features with a non-negligible positive or negative correlation with the response are relevant to the predictive performance of the models.
\added{
  In addition, other variables, such as \texttt{gwps\_gauss\_2\_2\_20}, \texttt{cs\_gauss\_2\_2\_20}, \texttt{gwps\_gauss\_2\_2\_55}, \texttt{cs\_gauss\_2\_2\_55}, \texttt{f\_07}, \texttt{sph}, \texttt{h\_cs\_20}, \texttt{teff}, and \texttt{length} have some relevance to the training of the models.
  In the specific case of the central periods of the second Gaussian function fitted in the GWPS and CS, although they are not the most relevant variables when training the models, the degree of importance assigned by the XGBoost algorithm to them can be attributed to the fact that the ground truth values were extracted from the central periods of the first fitted Gaussian functions, in conjunction with the algorithm's inherent ability to identify harmonic sequences.
  The 20 and 55-day filters are more relevant than the 80-day ones, given that K and M stars typically exhibit relatively slow rotation periods and the filter at 80-day may potentially be susceptible to instrumental issues.
}

For the most important predictors, we used tree-based approaches to build our models, which by their nature do not suffer from the inclusion of highly correlated features.
However, interpretability tools, such as importance estimations like the plots of \cref{fig:importance}, are hampered by collinearity and multicollinearity.
If two features are perfectly correlated, such as \replaced{\texttt{cs\_gauss\_3\_1\_20} and \texttt{cs\_gauss\_1\_1\_20}}{\texttt{cs\_gauss\_2\_1\_55} and \texttt{cs\_gauss\_3\_1\_55}}, they are likely to be selected by the algorithm.
In boosting, a link between a predictor and the response will remain stable once it has been learned by the model, and so the algorithm will stick to one of the correlated variables (but not both).
This can be seen for the pair of correlated variables above and others, such as \texttt{cs\_gauss\_1\_1\_20} and \replaced{\texttt{cs\_gauss\_1\_1\_55}}{\texttt{h\_cs\_20}} \added{or \texttt{sph\_cs\_err\_80}}, or \replaced{\texttt{gwps\_gauss\_2\_2\_55}}{\texttt{gwps\_gauss\_2\_2\_80}} and \replaced{\texttt{gwps\_gauss\_3\_3\_20}}{\texttt{gwps\_gauss\_3\_2\_80}} (just to name a few).
We end up realising that one of the variables plays an important role in generating predictions, but we do not realise that the other variable is also important in the link between the observations and the response unless we perform a correlation analysis of the features \citep{chen2018understand}, as we did in \cref{sec:stats}.
In this sense, any of the features identified as most important, in particular the CS and GWPS variables based on the \replaced{standard deviation}{central period} of the fitted Gaussian, could in principle be replaced by any other predictor highly correlated with them without significant loss of model performance.
In addition, as can be seen in the panels of \cref{fig:importance}, the importance profile for gradient boosting typically has a steep slope because the trees from boosting are interdependent and thus present correlated structures as the gradient evolves.
As a result, several predictors are selected across the trees, increasing their importance.

We believe that the level of importance attained by the CS and GWPS families of variables is related to the fact that most of the reference stellar rotation periods have been estimated from them.
As mentioned in \cref{sec:stats}, there is a strong correlation between some features of these families and the response.
One way to test this hypothesis would be to train new models from structured data obtained from synthetic light curves, produced by reliable stellar simulators, such as \added{the StarSim code \citep{rosich2020correcting} or} the PLATO Simulator \citep{marcos2014plato}.\added{
  \footnote{
    \added{PLATO stands for \textbf{PLA}netary \textbf{T}ransits and \textbf{O}scillations of the stars.}
  }
}
This might require the engineering of new variables, but in principle we would be able to control for the correlations between the predictors and the response.

The results indicate that the CS features are slightly more important for the model than those of the GWPS.
This is probably due to the fact that the CS is the product of the normalised GWPS and the ACF, thus amplifying peaks present in both and attenuating signals possibly due to instrumental effects, that manifest themselves differently in the GWPS and the ACF.

The most important TS variables are the activity proxies.
They are relevant in the sense that they are often in the top 10 of the most important variables.
It then becomes apparent that it is important to extract all possible activity proxies (photometric, magnetic, and others) directly from the light curves and to use them as predictors when building a model.

The weakest features for predicting stellar rotation periods are the astrophysical ones.
They \replaced{are rarely included in}{never make it into} the top 10 most important variables\replaced{.}{, and only}
\added{The exceptions are} the mass and the effective temperature\added{, which} sometimes make it into the top 30 \added{and were somewhat relevant to train DS0, despite the use of rotation periods as explanatory variables}.
This is consistent with the fact that it is not possible to predict stellar rotation periods from astrophysical variables alone.
However, the FliPer metrics are strongly correlated with some of the CS and GWPS predictors, as shown in \cref{sec:stats}.
Therefore, it is expected that the Flicker in Power metric, which is a proxy for the total power spectral density of a star, can play an important role in a regression ML model based on gradient boosting.
This means that the FliPer metrics could in principle replace any of the CS or GWPS features with which they are strongly correlated without significant loss of performance of the model.
The parameter space may also play a role in the small contribution of the Astro predictors.
We cannot exclude the possibility that examining a wider parameter space might change the picture, although we usually find a correlation in a narrower parameter space and not in a wider one.

\added{
  In selecting the most important predictors common to all data sets from DS1 to DS4, it is possible that some variables relevant to DS5 and DS6 may have been overlooked.
  An example is the mass, which was not selected for the final set of independent variables because it was not within the top 20 most important predictors for DS3 and DS4.
  Therefore, perhaps a more prudent approach, analogous to the methodology employed to generate DS2, should be used when identifying the most relevant predictors. 

  In \cref{fig:acc-iter2}, DS0 naturally stands out from the other models because it includes rotation periods in its explanatory variables.
  This fact is also responsible for the considerable discrepancy in the number of iterations required to achieve the optimal model observed between the training of DS0 and that of the remaining models.
}%
Reducing the number of predictors favours the XGB models, \added{especially for lower error tolerance regimes,} as can be seen from the fact that DS7 performs better overall than \deleted{DS0,} DS1 and DS2\added{, as well as the enhanced performance in terms of interval-based accuracy of DS3, DS4, DS5, and DS6 in comparison to all other models}.\added{\footnote{\added{
  We ran several tests, where we removed the features labelled ``\texttt{prot\_}'', but kept the central periods of the first Gaussian functions fitted to the CS and GWPS;
  we found that, in these cases, DS7 outperformed DS0.
  In fact, when the rotation period information was retained in the predictors, each of DS5, DS6, and DS7 performed better than DS0.
}}}
In particular, DS7 is on average less wrong, \added{and} has comparable accuracy, \deleted{smaller} testing RMSE and MAE, and \deleted{a larger} adjusted R$^2$ \replaced{to}{than} the other models.
\replaced{If anything, it has the same tendency}{It is also less likely} to overfit, and has less variance than \deleted{DS0,} DS1 and DS2 for slower rotators.
\added{%
  However, the selection of predictors must be made with great care to prevent eliminating variables that are important for training the model.
  Our approach of selecting the 20 most important variables common to the initial four models might have been less optimal, given the lack of significant performance improvement observed in DS6 relative to DS4.
}

The differences between the training and testing RMSE and MAE, although natural, indicate a degree of overfitting during the training process of the models built from the DS0 to DS7 data sets.
\deleted{The degree of overfitting increased slightly during iteration 2, but this did not seem to affect the quality performance of the models.}%
The differences between RMSE and MAE vary between the models generated with the different data sets, indicating dissimilar levels of variation of the individual errors.
However, these differences are not significant, even in the case of DS2, where they reached the highest value --- this seems to be related to the initial selection of predictors.

Prior to rotation period filtering, the models struggled to predict some of the very short rotation periods, typically less than a few days, and periods greater than \qty{45}{\day}.
These discrepancies are evident in the scatter plots of the models trained on the DS0 to DS6 data sets, and are further highlighted in \cref{fig:ratios}.
The over- and under-predictions are larger for very short rotation periods, below seven days.
Therefore, cases corresponding to K and M stars with rotation periods below \qty{7}{\day} or above \qty{45}{\day} are not suitable for training a predictive model using \textit{Kepler} data.
This is mainly due to the fact that
\begin{enumerate*}[label=(\alph*), font=\itshape]
  \item \textit{Kepler}'s quarters are 90 days long,
  \item at least two full cycles are required to obtain reliable observations,
  \item the process of stitching two or more \textit{Kepler} time series is not trivial and error-free, and
  \item signals from stars with rotation periods of less than seven days can be mimicked by fluxes from close binaries\replaced{,}{ or} classical pulsators\added{, and other sources of confusion}.
\end{enumerate*}
After the filtering of the rotation periods, the predictive power of the XGB models improved, as shown by the analysis of the models learned from DS3 to DS6.
By restricting the predictors to the set of the \replaced{28}{34} most important variables identified by the XGB models built with DS0 to DS4, and by filtering in stars with rotation periods between 7 and \qty{45}{\day}, we were able to train computationally cheap but solid models, with good predictive performance.

While the final models are characterised by a goodness of fit around \replaced{\qty{96}{\percent}}{\qty{98}{\percent}}, there are still a few outliers with predictions varying from approximately from 50 to \qty{200}{\percent} of the actual values.
In principle, increasing the number of training cases, controlling the errors in the measurements, and, most importantly, having a uniform distribution of training rotation periods (only possible with simulated data) would help to circumvent this problem and to improve the predictive power of the models.
In addition, we could perform feature engineering on the original light curves to try to get a better set of predictors.

Finally, our results are in line with those of \cite{breton2021rooster}.
\added{Taking into account the rotation periods in the explanatory variables, if we} \replaced{focus}{Focusing} on the \qty{10}{\percent} interval-accuracy metric, the classifier created with B21 achieved comparable results\replaced{.}{,}
\replaced{However,}{but} we have to take into account the fact that our accuracy metric is an approximation to the counterpart used to assess the quality of a classification model.
In addition, we did not visually check the results, nor did we change them after testing the model.
An important aspect to highlight is the fact that we used the \textit{hold-out method}, \ie we used an unseen-before testing set to assess the quality of the model.
This testing set was not part of the training of the model, nor dit it participate in the CV performed to optimise the hyperparameters of the models.
When assessing the quality of a model on a testing set, the predictive performance is typically more pessimistic than that obtained during the training with CV \citep{hastie2009elements,torgo2011data}.
The accuracies claimed by \cite{breton2021rooster} appear to be obtained with a process that is similar to a 2-fold CV with 100 repetitions, performed on the whole data set.
Therefore, even considering that their method is protected by the fact that they use the values of the target variable as predictors, we would expect the performance of their classifier to decrease slightly if the quality of the model were assessed on an unseen-before testing set.

\section{Conclusions}
\label{sec:conclusions}
In this paper we presented a novel method, based on a machine learning approach, to calculate stellar rotation periods of thousands of stars. 
We employed a regression analysis that makes use of tabular data extracted from light curves and that is based on the XGBoost algorithm.
Recently, attempts have been made to apply ANN to light curves and classification RFs to tabular data to make such predictions.
On the one hand, the former typically requires high computational resources;
on the other hand, with respect to the latter, we argue that
\begin{enumerate*}[label=(\alph*), font=\itshape]
  \item a classifier may not be the best approach to predict the rotation period, because it is a continuous variable, and
  \item using rotation periods as predictors may weaken the predictive performance of the model in the presence of unseen testing data, which do not contain rotation period values and have not been used to train the model.
\end{enumerate*}

Our main objective was to build robust and efficient ML models for automatically predicting stellar rotation periods of K and M stars from the \textit{Kepler} catalogue.
To this end, we applied the regression XGBoost algorithm to train models from seven data sets built from the data originally published by \cite{santos2019surface} and \cite{breton2021rooster}.

Given that \cite{breton2021rooster} built a classifier with their data set (B21), we developed an interval-based error and a metric, the interval-based accuracy, that allowed us to convert predicted rotation periods into a proportion of successful events.
We used this metric as a way to directly compare the results obtained with the two approaches (regression \vs classification). 

Initially, we used all the stars available in the base training set, \ie no stars were filtered out of the data set for model training.
Overall, the results show that the predictive performance of the models trained on DS0, a data set equivalent to B21, is comparable to the performance of the classifier trained on the latter, if we \added{keep rotation periods as independent variables and} are willing to accept a \qty{10}{\percent} error in the predictions.
\added{If we remove the rotation periods from the predictors, we get a \qty{10}{\percent}-accuracy of about \qty{90}{\percent}.}
The goodness of fit, as measured by the mean absolute value of the relative residuals, $\mu_\text{err}$, and by the adjusted coefficient of determination, $R^2_\text{adj}$, indicates that our models are robust, able to explain most of the variability in the response (approximately \replaced{\qty{96}{\percent})}{\qty{97}{\percent}}, and are on average wrong no more than \replaced{\qty{8}{\percent}}{\qty{6}{\percent}} of the time.

When we filter out stars with rotation periods below \qty{7}{\day} and above \qty{45}{\day}, the overall performance of the models increases, with about \replaced{\qtyrange{94}{96}{\percent}}{\qtyrange{96}{98}{\percent}} of the variability of the response explained by the predictors, and the \qty{10}{\percent} interval-accuracy equal to about \replaced{\qty{96}{\percent}}{\qty{95}{\percent}}.
In this case, the models were on average wrong about \replaced{\qtyrange{4}{6}{\percent}}{\qty{2}{\percent}} of the time.

\added{The mean relative uncertainty, as indicated by the extent of the error bars, was estimated to be approximately \qty{1.2}{\percent} for all the models.}
There is also an increase in the quality of the XGB models when the number of predictors is reduced through the selection of the most important variables that contribute to the learning process.

In view of the results reported in \cref{sec:results}, we conclude that the variables belonging to the CS and GWPS families are the most important for training a reliable XGBoost model with good predictive performance.
TS-type features, \ie those related to the structure of the light curve, such as $S_\text{ph}$, also have some relevance for estimating stellar rotation periods.
Although Astro-type variables, corresponding to astronomical observables and derived variables, do not have a significant weight in the importance plots, the FliPer metric and hence the power spectral density of a star can play a relevant role in model training, as they are strongly correlated with some of the variables in the CS and GWPS families.

By selecting the most important features, as measured during the training of the models obtained with the data sets DS0 to DS4, and excluding stars with rotation periods below \qty{7}{\day} and above \qty{45}{\day}, due to the characteristics of the \textit{Kepler} space observatory and astrophysical constraints, we were able to build an optimal subset of predictors from the set of available explanatory variables, from which robust regression ML models, with good predictive performance, could be trained.
Our results clearly show that XGBoost models trained on these reduced size data sets have comparable\deleted{performance to B21 and} \added{or} better performance than the models we previously obtained with the data sets \replaced{DS1}{DS0} to DS4.
We claim a reduction of about 140 predictors when compared to the largest data set (DS0).
The most important predictors, as measured by the \xgb\ \rlang\ package, are the\deleted{central periods and} standard deviations of the first Gaussian function\added{s} fitted to the CS and GWPS, followed by the photometric index proxy, $S_\text{ph}$.
By using a data set consisting mainly of these variables, it is possible to train XGB models with a performance comparable to or better than those built on much larger data sets.
By reducing the size of the sets by this order of magnitude, we are able to significantly improve the training time.
With the resources we had available, \ie shared machines running multiple processes simultaneously, we were able to reduce the time to build a model by several hours.

\added{
  In conclusion, the enhancement of data quality plays a pivotal role in the generation of accurate predictions using an XGBoost model.
  The reduction of the number of predictor variables while retaining the most important ones for model training represents a significant key contribution.

  The current limitations of our methodology are primarily due to the lack of training of the models to deal with false positives.
  Given the numerous sources of confusion, including binaries, classical pulsators, red giants signals, and even misclassified stars, we intend to apply this methodology to simulated data.
  This will enable us to consider all the known factors that could result in false detections and to train the models to address them.
}
%%%%%%%%%%%%%%%%%%%%%%%%%%%%%%%%%%%%%%%%%%%%%%%%%%%%%%%%%%%%%%%%%%%%%%%%%%%%%%%%

%ACKNOWLEDGEMETS%%%%%%%%%%%%%%%%%%%%%%%%%%%%%%%%%%%%%%%%%%%%%%%%%%%%%%%%%%%%%%%%
\begin{acknowledgements}
  \added{Nuno R. C. Gomes acknowledges support from the Atlantic Fisheries Fund research grant AFF-NS-1544 (Canada).}
  Fabio Del Sordo acknowledges support from a Marie Curie Action of the European Union (Grant agreement 101030103) and ``María de Maeztu'' award to the Institut de Ciències de l'Espai (CEX2020-001058-M).
  \added{We would like to express our gratitude to the reviewer for their perceptive comments and recommendations, which have contributed to the enhancement of the quality of this manuscript.}
  We gratefully acknowledge the CINECA award under the ISCRA initiative for the availability of high-performance computing resources and support, Sergio Messina and Nuccio Lanza for useful discussions, and Eva Ntormousi for comments on the manuscript.
  Nuno Gomes and Fabio Del Sordo would like to pay tribute to Luís Torgo, who left us far too soon.
\end{acknowledgements}


%BIBLIOGRAPHY%%%%%%%%%%%%%%%%%%%%%%%%%%%%%%%%%%%%%%%%%%%%%%%%%%%%%%%%%%%%%%%%%%%
\bibliographystyle{aa}
\bibliography{refs}
%%%%%%%%%%%%%%%%%%%%%%%%%%%%%%%%%%%%%%%%%%%%%%%%%%%%%%%%%%%%%%%%%%%%%%%%%%%%%%%%

%APPENDIX%%%%%%%%%%%%%%%%%%%%%%%%%%%%%%%%%%%%%%%%%%%%%%%%%%%%%%%%%%%%%%%%%%%%%%%
\begin{appendix}
\section{Plots}
\label{sec:plots}
In this appendix, we present a number of additional graphs to complement the information presented in the main body of the paper.
All figures were previously referenced, except those relating to elapsed times.

\begin{figure}[ht]
  \centering
  \includegraphics[width=0.45\textwidth]{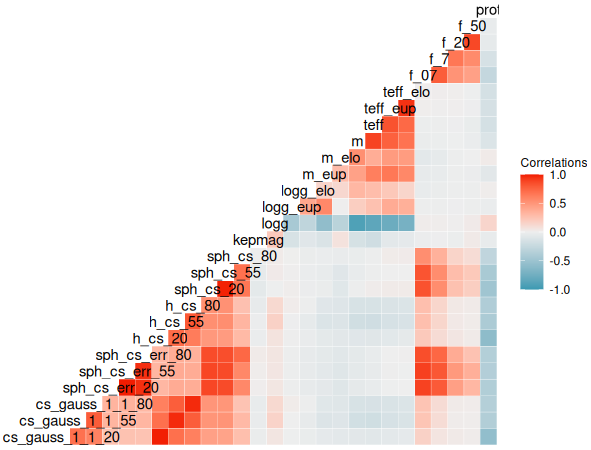}
  \caption{
    Correlations between the Astro variables, some of the CS predictors, and the response.
    The $F_{0.7}$ and $F_7$ FliPer metrics are highly correlated with some CS variables.
    }
  \label{fig:corr-astro-cs}
\end{figure}

\begin{figure}[ht]
  \centering
  \includegraphics[width=0.45\textwidth]{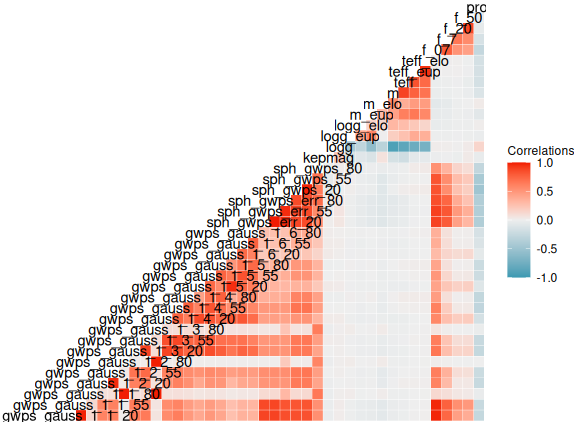}
  \caption{
    Correlations between the Astro variables, some of the GWPS predictors, and the response.
    Similarly to \cref{fig:corr-astro-cs}, the $F_{0.7}$ and $F_7$ FliPer metrics are highly correlated with some GWPS variables.
  }
  \label{fig:corr-astro-gwps}
\end{figure}

\Cref{fig:corr-astro-cs} shows the correlations between predictors belonging to the Astro and CS families of variables, and the response.
Variables that are highly correlated with each other, such as \texttt{sph\_cs\_20} and \texttt{f\_07}, can in principle be swapped in the model without loss of performance.
The situation is similar for the predictors belonging to the GWPS family of variables, whose correlations are shown in \cref{fig:corr-astro-gwps}.
Only negative or weak correlations with the response are shown in these plots.
However, features that are strongly correlated with the target variable would be highlighted in full plots of the CS and GWPS predictors and the target variable.


\begin{figure}[!ht]
  \centering
  \includegraphics[width=\linewidth]{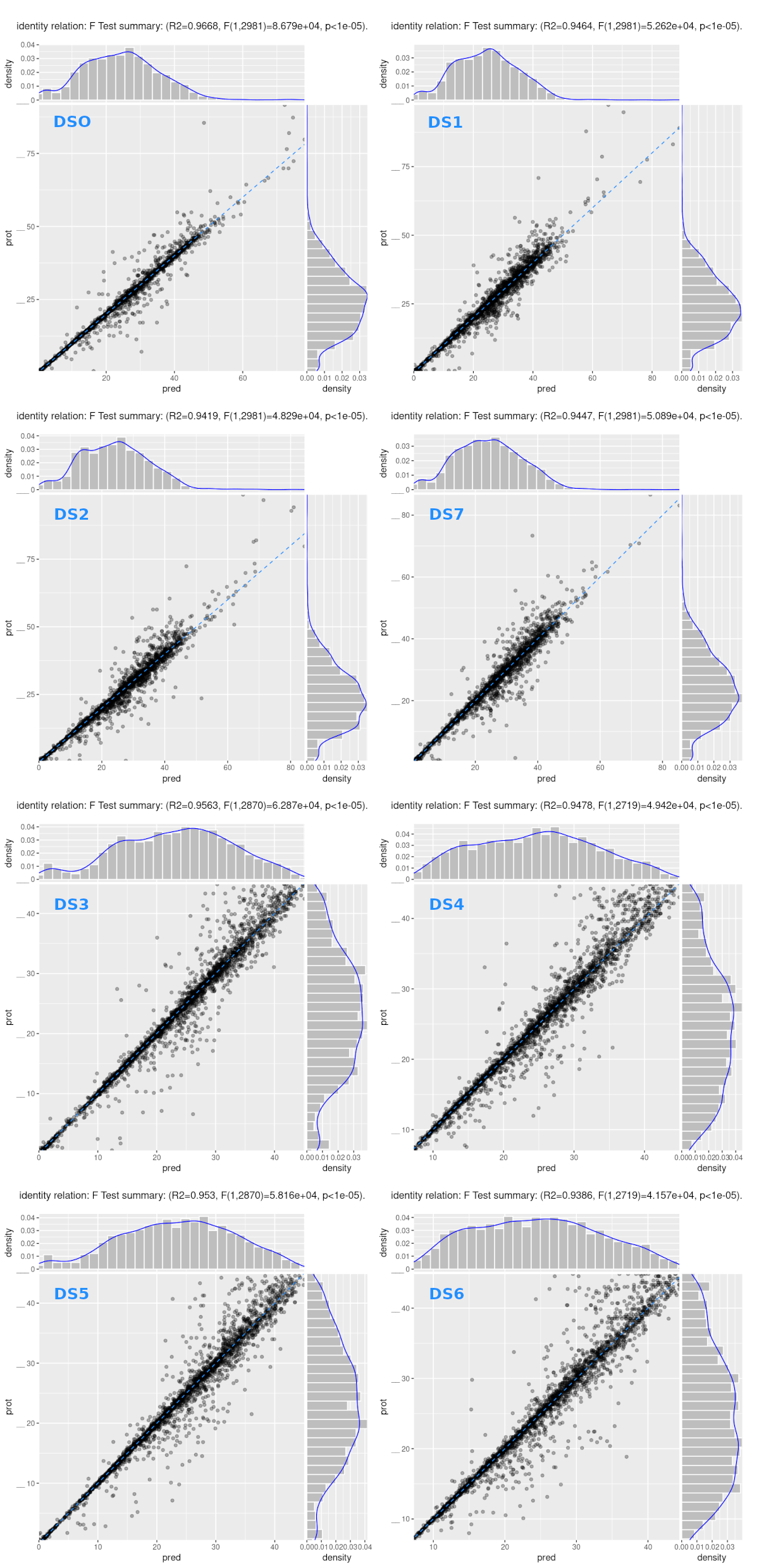}
  \vspace{1mm}
  \caption{
    Scatter and marginal density plots of the reference rotation period as a function of the predicted values for the models built during iteration 2.
    The blue dashed lines indicate the identity function.
    DS7 has been placed next to DS1 and DS2 to facilitate comparison of the models.
  }
  \label{fig:scatter-density-grid}
\end{figure}

\Cref{fig:scatter-density-grid} shows the scatter plots of the ground truth \vs the predictions for the models generated during iteration 2, including the marginal density plots of the predictions and the reference rotation periods.
Similar to \cref{fig:scatter-grid}, the blue dashed line indicates the identity function.
Each panel reports the goodness of fit and significance of the relationship, using $R^2$, the F-test and the corresponding $p$-value, respectively.
Overall, the sets of predictors are jointly significant.

\begin{figure}[!ht]
  \centering
  \includegraphics[width=\linewidth]{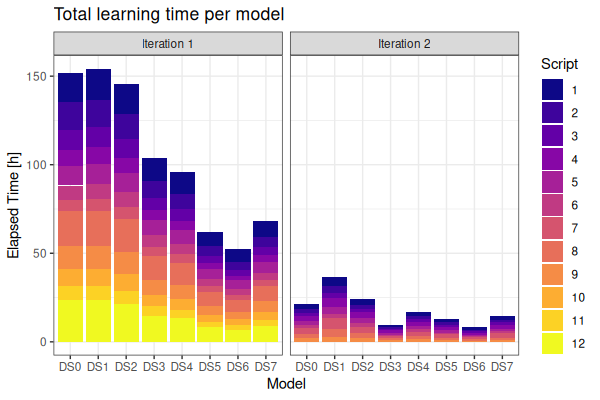}
  % \vspace{1mm}
  \caption{
    Total learning time per model in iterations 1 and 2.
    To build the models, the hyperparameter grids were split into 12 and nine sub-grids in iteration 1 and 2, respectively.
    The height of the scripts is proportional to the amount of time it takes to execute them.
  }
  \label{fig:total-learning-time}
\end{figure}

\begin{figure}[!ht]
  \centering
  \includegraphics[width=\linewidth]{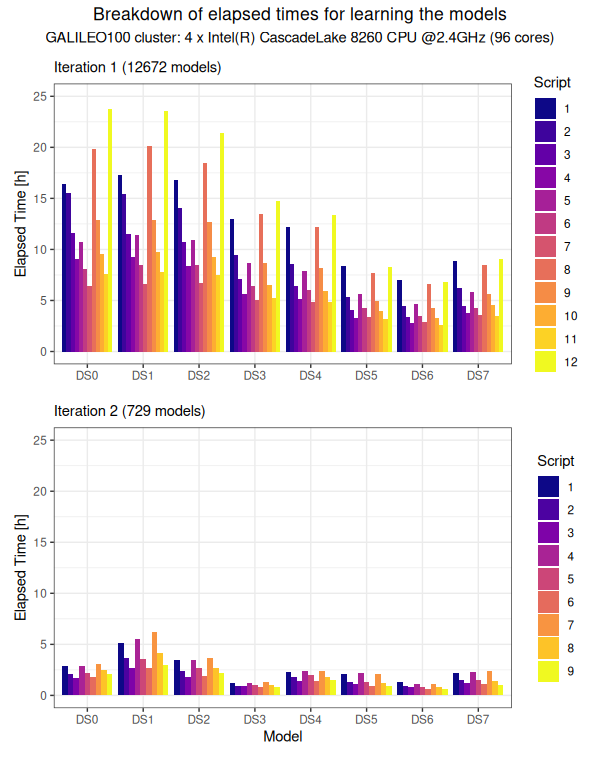}
  % \vspace{1mm}
  \caption{
    Breakdown of learning times for the models in iteration 1 and 2.
    Even though the total run time amounted to \qty{150}{\hour} in iteration 1 (see \cref{fig:total-learning-time}), the longest models took no more than \qty{24}{\hour} to complete.
  }
  \label{fig:breakdown-learning-time}
\end{figure}

The total learning time per model in iterations 1 and 2 is shown in \cref{fig:total-learning-time}.
In order to optimise the generation of the models, the hyperparameter grids were divided into several sub-grids (12 and nine in iteration 1 and 2, respectively), which were started in parallel.
In this way, the longest models, with more than \qty{150}{\hour} of learning time, took less than \qty{24}{\hour} to complete, as shown in \cref{fig:breakdown-learning-time}.
Using this approach, the fastest models in iteration 2 took less than two hours to train.


\vfill\null
\phantom{Column 2}

\end{appendix}
%%%%%%%%%%%%%%%%%%%%%%%%%%%%%%%%%%%%%%%%%%%%%%%%%%%%%%%%%%%%%%%%%%%%%%%%%%%%%%%%

\end{document}